\documentclass[art11]{article}
\usepackage{graphicx}
\usepackage{amsmath,amssymb}
\usepackage{caption}
\usepackage{times}
\usepackage{subfigure}
\usepackage{cite}
\usepackage{multirow}
\usepackage{longtable}
\usepackage[affil-sl] {authblk}
\usepackage[ruled,vlined]{algorithm2e}
\usepackage{fancyvrb}

\def\PsfigVersion{1.9}
\ifx\undefined\psfig\else \fi

\let\LaTeXAtSign=\@
\let\@=\relax
\edef\psfigRestoreAt{\catcode`\@=\number\catcode`@\relax}
\catcode`\@=11\relax
\newwrite\@unused
\def\ps@typeout#1{{\let\protect\string\immediate\write\@unused{#1}}}
\ps@typeout{psfig/tex \PsfigVersion}

\def\figurepath{./}

\def\@nnil{\@nil}
\def\@empty{}
\def\@psdonoop#1\@@#2#3{}
\def\@psdo#1:=#2\do#3{\edef\@psdotmp{#2}\ifx\@psdotmp\@empty \else
    \expandafter\@psdoloop#2,\@nil,\@nil\@@#1{#3}\fi}
\def\@psdoloop#1,#2,#3\@@#4#5{\def#4{#1}\ifx #4\@nnil \else
       #5\def#4{#2}\ifx #4\@nnil \else#5\@ipsdoloop #3\@@#4{#5}\fi\fi}
\def\@ipsdoloop#1,#2\@@#3#4{\def#3{#1}\ifx #3\@nnil 
       \let\@nextwhile=\@psdonoop \else
      #4\relax\let\@nextwhile=\@ipsdoloop\fi\@nextwhile#2\@@#3{#4}}
\def\@tpsdo#1:=#2\do#3{\xdef\@psdotmp{#2}\ifx\@psdotmp\@empty \else
    \@tpsdoloop#2\@nil\@nil\@@#1{#3}\fi}
\def\@tpsdoloop#1#2\@@#3#4{\def#3{#1}\ifx #3\@nnil 
       \let\@nextwhile=\@psdonoop \else
      #4\relax\let\@nextwhile=\@tpsdoloop\fi\@nextwhile#2\@@#3{#4}}
\ifx\undefined\fbox
\newdimen\fboxrule
\newdimen\fboxsep
\newdimen\ps@tempdima
\newbox\ps@tempboxa
\long\def\fbox#1{\leavevmode\setbox\ps@tempboxa\hbox{#1}\ps@tempdima\fboxrule
    \advance\ps@tempdima \fboxsep \advance\ps@tempdima \dp\ps@tempboxa
   \hbox{\lower \ps@tempdima\hbox
  {\vbox{\hrule height \fboxrule
          \hbox{\vrule width \fboxrule \hskip\fboxsep
          \vbox{\vskip\fboxsep \box\ps@tempboxa\vskip\fboxsep}\hskip 
                 \fboxsep\vrule width \fboxrule}
                 \hrule height \fboxrule}}}}
\fi
\newread\ps@stream
\newif\ifnot@eof       
\newif\if@noisy        
\newif\if@atend        
\newif\if@psfile       
{\catcode`\%=12\global\gdef\epsf@start{
\def\epsf@PS{PS}
\def\epsf@getbb#1{%
\openin\ps@stream=#1
\ifeof\ps@stream\ps@typeout{Error, File #1 not found}\else
   {\not@eoftrue \chardef\other=12
    \def\do##1{\catcode`##1=\other}\dospecials \catcode`\ =10
    \loop
       \if@psfile
	  \read\ps@stream to \epsf@fileline
       \else{
	  \obeyspaces
          \read\ps@stream to \epsf@tmp\global\let\epsf@fileline\epsf@tmp}
       \fi
       \ifeof\ps@stream\not@eoffalse\else
       \if@psfile\else
       \expandafter\epsf@test\epsf@fileline:. \\%
       \fi
          \expandafter\epsf@aux\epsf@fileline:. \\%
       \fi
   \ifnot@eof\repeat
   }\closein\ps@stream\fi}%
\long\def\epsf@test#1#2#3:#4\\{\def\epsf@testit{#1#2}
			\ifx\epsf@testit\epsf@start\else
\ps@typeout{Warning! File does not start with `\epsf@start'.  It may not be a PostScript file.}
			\fi
			\@psfiletrue} 
{\catcode`\%=12\global\let\epsf@percent=
\long\def\epsf@aux#1#2:#3\\{\ifx#1\epsf@percent
   \def\epsf@testit{#2}\ifx\epsf@testit\epsf@bblit
	\@atendfalse
        \epsf@atend #3 . \\%
	\if@atend	
	   \if@verbose{
		\ps@typeout{psfig: found `(atend)'; continuing search}
	   }\fi
        \else
        \epsf@grab #3 . . . \\%
        \not@eoffalse
        \global\no@bbfalse
        \fi
   \fi\fi}%
\def\epsf@grab #1 #2 #3 #4 #5\\{%
   \global\def\epsf@llx{#1}\ifx\epsf@llx\empty
      \epsf@grab #2 #3 #4 #5 .\\\else
   \global\def\epsf@lly{#2}%
   \global\def\epsf@urx{#3}\global\def\epsf@ury{#4}\fi}%
\def\epsf@atendlit{(atend)} 
\def\epsf@atend #1 #2 #3\\{%
   \def\epsf@tmp{#1}\ifx\epsf@tmp\empty
      \epsf@atend #2 #3 .\\\else
   \ifx\epsf@tmp\epsf@atendlit\@atendtrue\fi\fi}

\chardef\psletter = 11 
\chardef\other = 12

\newif \ifdebug 
\newif\ifc@mpute 
\c@mputetrue 

\let\then = \relax
\def\r@dian{pt }
\let\r@dians = \r@dian
\let\dimensionless@nit = \r@dian
\let\dimensionless@nits = \dimensionless@nit
\def\internal@nit{sp }
\let\internal@nits = \internal@nit
\newif\ifstillc@nverging
\def \Mess@ge #1{\ifdebug \then \message {#1} \fi}

{ 
	\catcode `\@ = \psletter
	\gdef \nodimen {\expandafter \n@dimen \the \dimen}
	\gdef \term #1 #2 #3%
	       {\edef \t@ {\the #1}
		\edef \t@@ {\expandafter \n@dimen \the #2\r@dian}%
		\t@rm {\t@} {\t@@} {#3}%
	       }
	\gdef \t@rm #1 #2 #3%
	       {{%
		\count 0 = 0
		\dimen 0 = 1 \dimensionless@nit
		\dimen 2 = #2\relax
		\Mess@ge {Calculating term #1 of \nodimen 2}%
		\loop
		\ifnum	\count 0 < #1
		\then	\advance \count 0 by 1
			\Mess@ge {Iteration \the \count 0 \space}%
			\Multiply \dimen 0 by {\dimen 2}%
			\Mess@ge {After multiplication, term = \nodimen 0}%
			\Divide \dimen 0 by {\count 0}%
			\Mess@ge {After division, term = \nodimen 0}%
		\repeat
		\Mess@ge {Final value for term #1 of 
				\nodimen 2 \space is \nodimen 0}%
		\xdef \Term {#3 = \nodimen 0 \r@dians}%
		\aftergroup \Term
	       }}
	\catcode `\p = \other
	\catcode `\t = \other
	\gdef \n@dimen #1pt{#1} 
}

\def \Divide #1by #2{\divide #1 by #2} 

\def \Multiply #1by #2
       {{
	\count 0 = #1\relax
	\count 2 = #2\relax
	\count 4 = 65536
	\Mess@ge {Before scaling, count 0 = \the \count 0 \space and
			count 2 = \the \count 2}%
	\ifnum	\count 0 > 32767 
	\then	\divide \count 0 by 4
		\divide \count 4 by 4
	\else	\ifnum	\count 0 < -32767
		\then	\divide \count 0 by 4
			\divide \count 4 by 4
		\else
		\fi
	\fi
	\ifnum	\count 2 > 32767 
	\then	\divide \count 2 by 4
		\divide \count 4 by 4
	\else	\ifnum	\count 2 < -32767
		\then	\divide \count 2 by 4
			\divide \count 4 by 4
		\else
		\fi
	\fi
	\multiply \count 0 by \count 2
	\divide \count 0 by \count 4
	\xdef \product {#1 = \the \count 0 \internal@nits}%
	\aftergroup \product
       }}

\def\r@duce{\ifdim\dimen0 > 90\r@dian \then   
		\multiply\dimen0 by -1
		\advance\dimen0 by 180\r@dian
		\r@duce
	    \else \ifdim\dimen0 < -90\r@dian \then  
		\advance\dimen0 by 360\r@dian
		\r@duce
		\fi
	    \fi}

\def\Sine#1%
       {{%
	\dimen 0 = #1 \r@dian
	\r@duce
	\ifdim\dimen0 = -90\r@dian \then
	   \dimen4 = -1\r@dian
	   \c@mputefalse
	\fi
	\ifdim\dimen0 = 90\r@dian \then
	   \dimen4 = 1\r@dian
	   \c@mputefalse
	\fi
	\ifdim\dimen0 = 0\r@dian \then
	   \dimen4 = 0\r@dian
	   \c@mputefalse
	\fi
	\ifc@mpute \then
		\divide\dimen0 by 180
		\dimen0=3.141592654\dimen0
		\dimen 2 = 3.1415926535897963\r@dian 
		\divide\dimen 2 by 2 
		\Mess@ge {Sin: calculating Sin of \nodimen 0}%
		\count 0 = 1 
		\dimen 2 = 1 \r@dian 
		\dimen 4 = 0 \r@dian 
		\loop
			\ifnum	\dimen 2 = 0 
			\then	\stillc@nvergingfalse 
			\else	\stillc@nvergingtrue
			\fi
			\ifstillc@nverging 
			\then	\term {\count 0} {\dimen 0} {\dimen 2}%
				\advance \count 0 by 2
				\count 2 = \count 0
				\divide \count 2 by 2
				\ifodd	\count 2 
				\then	\advance \dimen 4 by \dimen 2
				\else	\advance \dimen 4 by -\dimen 2
				\fi
		\repeat
	\fi		
			\xdef \sine {\nodimen 4}%
       }}

\def\Cosine#1{\ifx\sine\UnDefined\edef\Savesine{\relax}\else
		             \edef\Savesine{\sine}\fi
	{\dimen0=#1\r@dian\advance\dimen0 by 90\r@dian
	 \Sine{\nodimen 0}
	 \xdef\cosine{\sine}
	 \xdef\sine{\Savesine}}}	      

\def\psdraft{
	\def\@psdraft{0}
}
\def\psfull{
	\def\@psdraft{100}
}

\psfull

\newif\if@scalefirst
\def\psscalefirst{\@scalefirsttrue}
\def\psrotatefirst{\@scalefirstfalse}
\psrotatefirst

\newif\if@draftbox
\def\psnodraftbox{
	\@draftboxfalse
}
\def\psdraftbox{
	\@draftboxtrue
}
\@draftboxtrue

\newif\if@prologfile
\newif\if@postlogfile
\def\pssilent{
	\@noisyfalse
}
\def\psnoisy{
	\@noisytrue
}
\psnoisy
\newif\if@bbllx
\newif\if@bblly
\newif\if@bburx
\newif\if@bbury
\newif\if@height
\newif\if@width
\newif\if@rheight
\newif\if@rwidth
\newif\if@angle
\newif\if@clip
\newif\if@verbose
\def\@p@@sclip#1{\@cliptrue}

\newif\if@decmpr

\def\@p@@sfigure#1{\def\@p@sfile{null}\def\@p@sbbfile{null}
	        \openin1=#1.bb
		\ifeof1\closein1
	        	\openin1=\figurepath#1.bb
			\ifeof1\closein1
			        \openin1=#1
				\ifeof1\closein1%
				       \openin1=\figurepath#1
					\ifeof1
					   \ps@typeout{Error, File #1 not found}
						\if@bbllx\if@bblly
				   		\if@bburx\if@bbury
			      				\def\@p@sfile{#1}%
			      				\def\@p@sbbfile{#1}%
							\@decmprfalse
				  	   	\fi\fi\fi\fi
					\else\closein1
				    		\def\@p@sfile{\figurepath#1}%
				    		\def\@p@sbbfile{\figurepath#1}%
						\@decmprfalse
	                       		\fi%
			 	\else\closein1%
					\def\@p@sfile{#1}
					\def\@p@sbbfile{#1}
					\@decmprfalse
			 	\fi
			\else
				\def\@p@sfile{\figurepath#1}
				\def\@p@sbbfile{\figurepath#1.bb}
				\@decmprtrue
			\fi
		\else
			\def\@p@sfile{#1}
			\def\@p@sbbfile{#1.bb}
			\@decmprtrue
		\fi}

\def\@p@@sfile#1{\@p@@sfigure{#1}}

\def\@p@@sbbllx#1{
		\@bbllxtrue
		\dimen100=#1
		\edef\@p@sbbllx{\number\dimen100}
}
\def\@p@@sbblly#1{
		\@bbllytrue
		\dimen100=#1
		\edef\@p@sbblly{\number\dimen100}
}
\def\@p@@sbburx#1{
		\@bburxtrue
		\dimen100=#1
		\edef\@p@sbburx{\number\dimen100}
}
\def\@p@@sbbury#1{
		\@bburytrue
		\dimen100=#1
		\edef\@p@sbbury{\number\dimen100}
}
\def\@p@@sheight#1{
		\@heighttrue
		\dimen100=#1
   		\edef\@p@sheight{\number\dimen100}
}
\def\@p@@swidth#1{
		\@widthtrue
		\dimen100=#1
		\edef\@p@swidth{\number\dimen100}
}
\def\@p@@srheight#1{
		\@rheighttrue
		\dimen100=#1
		\edef\@p@srheight{\number\dimen100}
}
\def\@p@@srwidth#1{
		\@rwidthtrue
		\dimen100=#1
		\edef\@p@srwidth{\number\dimen100}
}
\def\@p@@sangle#1{
		\@angletrue
		\edef\@p@sangle{#1} 
}
\def\@p@@ssilent#1{ 
		\@verbosefalse
}
\def\@p@@sprolog#1{\@prologfiletrue\def\@prologfileval{#1}}
\def\@p@@spostlog#1{\@postlogfiletrue\def\@postlogfileval{#1}}
\def\@cs@name#1{\csname #1\endcsname}
\def\@setparms#1=#2,{\@cs@name{@p@@s#1}{#2}}
\def\ps@init@parms{
		\@bbllxfalse \@bbllyfalse
		\@bburxfalse \@bburyfalse
		\@heightfalse \@widthfalse
		\@rheightfalse \@rwidthfalse
		\def\@p@sbbllx{}\def\@p@sbblly{}
		\def\@p@sbburx{}\def\@p@sbbury{}
		\def\@p@sheight{}\def\@p@swidth{}
		\def\@p@srheight{}\def\@p@srwidth{}
		\def\@p@sangle{0}
		\def\@p@sfile{} \def\@p@sbbfile{}
		\def\@p@scost{10}
		\def\@sc{}
		\@prologfilefalse
		\@postlogfilefalse
		\@clipfalse
		\if@noisy
			\@verbosetrue
		\else
			\@verbosefalse
		\fi
}
\def\parse@ps@parms#1{
	 	\@psdo\@psfiga:=#1\do
		   {\expandafter\@setparms\@psfiga,}}
\newif\ifno@bb
\def\bb@missing{
	\if@verbose{
		\ps@typeout{psfig: searching \@p@sbbfile \space  for bounding box}
	}\fi
	\no@bbtrue
	\epsf@getbb{\@p@sbbfile}
        \ifno@bb \else \bb@cull\epsf@llx\epsf@lly\epsf@urx\epsf@ury\fi
}	
\def\bb@cull#1#2#3#4{
	\dimen100=#1 bp\edef\@p@sbbllx{\number\dimen100}
	\dimen100=#2 bp\edef\@p@sbblly{\number\dimen100}
	\dimen100=#3 bp\edef\@p@sbburx{\number\dimen100}
	\dimen100=#4 bp\edef\@p@sbbury{\number\dimen100}
	\no@bbfalse
}
\newdimen\p@intvaluex
\newdimen\p@intvaluey
\def\rotate@#1#2{{\dimen0=#1 sp\dimen1=#2 sp
		  \global\p@intvaluex=\cosine\dimen0
		  \dimen3=\sine\dimen1
		  \global\advance\p@intvaluex by -\dimen3
		  \global\p@intvaluey=\sine\dimen0
		  \dimen3=\cosine\dimen1
		  \global\advance\p@intvaluey by \dimen3
		  }}
\def\compute@bb{
		\no@bbfalse
		\if@bbllx \else \no@bbtrue \fi
		\if@bblly \else \no@bbtrue \fi
		\if@bburx \else \no@bbtrue \fi
		\if@bbury \else \no@bbtrue \fi
		\ifno@bb \bb@missing \fi
		\ifno@bb \ps@typeout{FATAL ERROR: no bb supplied or found}
			\no-bb-error
		\fi
		\count203=\@p@sbburx
		\count204=\@p@sbbury
		\advance\count203 by -\@p@sbbllx
		\advance\count204 by -\@p@sbblly
		\edef\ps@bbw{\number\count203}
		\edef\ps@bbh{\number\count204}
		\if@angle 
			\Sine{\@p@sangle}\Cosine{\@p@sangle}
	        	{\dimen100=\maxdimen\xdef\r@p@sbbllx{\number\dimen100}
					    \xdef\r@p@sbblly{\number\dimen100}
			                    \xdef\r@p@sbburx{-\number\dimen100}
					    \xdef\r@p@sbbury{-\number\dimen100}}
                        \def\minmaxtest{
			   \ifnum\number\p@intvaluex<\r@p@sbbllx
			      \xdef\r@p@sbbllx{\number\p@intvaluex}\fi
			   \ifnum\number\p@intvaluex>\r@p@sbburx
			      \xdef\r@p@sbburx{\number\p@intvaluex}\fi
			   \ifnum\number\p@intvaluey<\r@p@sbblly
			      \xdef\r@p@sbblly{\number\p@intvaluey}\fi
			   \ifnum\number\p@intvaluey>\r@p@sbbury
			      \xdef\r@p@sbbury{\number\p@intvaluey}\fi
			   }
			\rotate@{\@p@sbbllx}{\@p@sbblly}
			\minmaxtest
			\rotate@{\@p@sbbllx}{\@p@sbbury}
			\minmaxtest
			\rotate@{\@p@sbburx}{\@p@sbblly}
			\minmaxtest
			\rotate@{\@p@sbburx}{\@p@sbbury}
			\minmaxtest
			\edef\@p@sbbllx{\r@p@sbbllx}\edef\@p@sbblly{\r@p@sbblly}
			\edef\@p@sbburx{\r@p@sbburx}\edef\@p@sbbury{\r@p@sbbury}
		\fi
		\count203=\@p@sbburx
		\count204=\@p@sbbury
		\advance\count203 by -\@p@sbbllx
		\advance\count204 by -\@p@sbblly
		\edef\@bbw{\number\count203}
		\edef\@bbh{\number\count204}
}
\def\in@hundreds#1#2#3{\count240=#2 \count241=#3
		     \count100=\count240	
		     \divide\count100 by \count241
		     \count101=\count100
		     \multiply\count101 by \count241
		     \advance\count240 by -\count101
		     \multiply\count240 by 10
		     \count101=\count240	
		     \divide\count101 by \count241
		     \count102=\count101
		     \multiply\count102 by \count241
		     \advance\count240 by -\count102
		     \multiply\count240 by 10
		     \count102=\count240	
		     \divide\count102 by \count241
		     \count200=#1\count205=0
		     \count201=\count200
			\multiply\count201 by \count100
		 	\advance\count205 by \count201
		     \count201=\count200
			\divide\count201 by 10
			\multiply\count201 by \count101
			\advance\count205 by \count201
		     \count201=\count200
			\divide\count201 by 100
			\multiply\count201 by \count102
			\advance\count205 by \count201
		     \edef\@result{\number\count205}
}
\def\compute@wfromh{
		\in@hundreds{\@p@sheight}{\@bbw}{\@bbh}
		\edef\@p@swidth{\@result}
}
\def\compute@hfromw{
	        \in@hundreds{\@p@swidth}{\@bbh}{\@bbw}
		\edef\@p@sheight{\@result}
}
\def\compute@handw{
		\if@height 
			\if@width
			\else
				\compute@wfromh
			\fi
		\else 
			\if@width
				\compute@hfromw
			\else
				\edef\@p@sheight{\@bbh}
				\edef\@p@swidth{\@bbw}
			\fi
		\fi
}
\def\compute@resv{
		\if@rheight \else \edef\@p@srheight{\@p@sheight} \fi
		\if@rwidth \else \edef\@p@srwidth{\@p@swidth} \fi
}
\def\compute@sizes{
	\compute@bb
	\if@scalefirst\if@angle
	\if@width
	   \in@hundreds{\@p@swidth}{\@bbw}{\ps@bbw}
	   \edef\@p@swidth{\@result}
	\fi
	\if@height
	   \in@hundreds{\@p@sheight}{\@bbh}{\ps@bbh}
	   \edef\@p@sheight{\@result}
	\fi
	\fi\fi
	\compute@handw
	\compute@resv}

\def\psfig#1{\vbox {
	\ps@init@parms
	\parse@ps@parms{#1}
	\compute@sizes
	\ifnum\@p@scost<\@psdraft{
		\special{ps::[begin] 	\@p@swidth \space \@p@sheight \space
				\@p@sbbllx \space \@p@sbblly \space
				\@p@sbburx \space \@p@sbbury \space
				startTexFig \space }
		\if@angle
			\special {ps:: \@p@sangle \space rotate \space} 
		\fi
		\if@clip{
			\if@verbose{
				\ps@typeout{(clip)}
			}\fi
			\special{ps:: doclip \space }
		}\fi
		\if@prologfile
		    \special{ps: plotfile \@prologfileval \space } \fi
		\if@decmpr{
			\if@verbose{
				\ps@typeout{psfig: including \@p@sfile.Z \space }
			}\fi
			\special{ps: plotfile "`zcat \@p@sfile.Z" \space }
		}\else{
			\if@verbose{
				\ps@typeout{psfig: including \@p@sfile \space }
			}\fi
			\special{ps: plotfile \@p@sfile \space }
		}\fi
		\if@postlogfile
		    \special{ps: plotfile \@postlogfileval \space } \fi
		\special{ps::[end] endTexFig \space }
		\vbox to \@p@srheight sp{
			\hbox to \@p@srwidth sp{
				\hss
			}
		\vss
		}
	}\else{
		\if@draftbox{		
			\hbox{\frame{\vbox to \@p@srheight sp{
			\vss
			\hbox to \@p@srwidth sp{ \hss \@p@sfile \hss }
			\vss
			}}}
		}\else{
			\vbox to \@p@srheight sp{
			\vss
			\hbox to \@p@srwidth sp{\hss}
			\vss
			}
		}\fi

	}\fi
}}
\psfigRestoreAt
\let\@=\LaTeXAtSign

\includeonly {}
\setlength{\topmargin}{0.0in}
\setlength{\oddsidemargin}{0.0in}
\setlength{\footskip}{0.6in}
\setlength{\headheight}{0.0in}
\setlength{\headsep}{0.0in}
\setlength{\textheight}{9.0in}
\setlength{\textwidth}{6.7in}
\setlength{\parindent}{0in}
\setlength{\parskip}{0.1in}
\setlength{\columnsep}{2.5pc}

\renewcommand{\baselinestretch}{1.0}

\begin {document}

\title{Dynamic Diagnosis of the Progress and Shortcomings of \\ Student Learning using Machine Learning based on \\ Cognitive, Social, and Emotional Features}

\author[]{
Alex Doboli$^1$, Simona Doboli$^2$, Ryan Duke$^1$, Sangjin Hong$^1$ and Wendy Tang$^1$
}

\maketitle

$^1$ Department of Electrical and Computer Engineering, Stony Brook University, Stony Brook NY 117954-2350, \\ Email: alex.doboli, ryan.duke, sangjin.hong, wendy.tang@stonybrook.edu \\%

$^2$ Department of Computer Science, Hofstra University, Hempstead NY 11549, Email: simona.doboli@hofstra.edu

\maketitle

\begin{abstract}

Student diversity, like academic background, learning styles, career and life goals, ethnicity, age, social and emotional characteristics, course load and work schedule, offers unique opportunities in education, like learning new skills, peer mentoring and example setting. But student diversity can be challenging too as it adds variability in the way in which students learn and progress over time. A single teaching approach is likely to be ineffective and result in students not meeting their potential. Automated support could address limitations of traditional teaching by continuously assessing student learning and implementing needed interventions. This paper discusses a novel methodology based on data analytics and Machine Learning to measure and causally diagnose the progress and shortcomings of student learning, and then utilizes the insight gained on individuals to optimize learning. Diagnosis pertains to dynamic diagnostic formative assessment, which aims to uncover the causes of learning shortcomings. The methodology groups learning difficulties into four categories: recall from memory, concept adjustment, concept modification, and problem decomposition into sub-goals (sub-problems) and concept combination. Data models are predicting the occurrence of each of the four challenge types, as well as a student's learning trajectory. The models can be used to automatically create real-time, student-specific interventions (e.g., learning cues) to address less understood concepts. We envision that the system will enable new adaptive pedagogical approaches to unleash student learning potential through customization of the course material to the background, abilities, situation, and progress of each student; and leveraging diversity-related learning experiences.

\end{abstract}



\thispagestyle{empty}

\section {Introduction}

Literature on education research shows that diversity offers unique opportunities in education, if everyone's uniqueness is understood and respected by the rest~\cite{Pociask2017, Shen2007, Stewart2018}. For example, fast learners do not feel hindered by slow learning students, because they understand how the entire class can benefit from people with different learning styles, diverse backgrounds, and from other social strata~\cite{Black1999, Chatman1998being, Shen2007, Stewart2018}. Student diversity exists along multiple cognitive, social, and cultural dimensions, i.e. academic background, learning styles with strengths and weaknesses, career and life goals and expectations, ethnicity, age, social and emotional characteristics, course load, work schedule, and so on. Differences can expose unexpected possibilities for learning new skills, like peer mentoring and example setting, and hence increase the awareness and appreciation for human respect and compassion~\cite{Black1999, Chatman1998being, Pociask2017, van2004work}. 



But student diversity can be challenging for teaching, as it adds significant variability in the way in which students learn and progress over time. A single teaching approach, especially for large student populations, is likely to be ineffective. Typical learning activities in engineering and computer science courses include lectures, laboratory, and individual or team-based assignments, like home works and course projects. The instructor usually teaches one lecture and then assigns one homework or one laboratory activity to the whole class. Instructors get feedback on individual's and class learning outcomes after grading assignments. In larger classes with no teaching assistants (TA) or limited TA support, grading typically takes at least one week. This often results in students receiving delayed feedback about learning and their progress, and in teachers making late adjustments on teaching more complex concepts. In this overwhelmingly common approach, students who need help receive it too late or too little, and advanced students may not be challenged enough. Given the diversity of students along multiple dimensions, the one-way approach to education results in undergraduates not meeting their education potential. This paper argues that automated support is critical in enabling educators to assess student learning and progress, as well as the needed interventions to maximize student outcomes.

Numerous educational technologies have been suggested to improve teaching and student learning \cite{Atif2013, Ekowo2016, Kassab2020, Li2018}. Research work has developed novel devices (like RFID-based attendance monitoring, remote labs, classroom management, etc.) \cite{Kassab2020, Sula2013}, student performance analysis \cite{Atif2013, Li2018}, learning resource suggestion \cite{Atif2013}, targeted student advising \cite{Attaran2018, Ekowo2016}, adaptive learning \cite{Ekowo2016}, course recommendation \cite{Ekowo2016}, campus management \cite{Kassab2020}, enrollment management \cite{Ekowo2016}. Another school of thought considers the insight gained by psychologists on the cognitive processes during learning \cite{Bell1996, Keene2010, Piaget2008}. Then, systems, like Cognitive Tutor, use this insight to suggest customized exercises in math or physics \cite{Ritter2007, Vincent2002}. Recent work suggests powerful data analytics and Machine Learning (ML) methods for teaching strategy analysis \cite{Atif2013, Li2018} and prediction of course outcomes, graduation rates, or student success \cite{Ekowo2016, Li2018, Sclater2016}. 

While some approaches consider personalizing the teaching material to student abilities, existing educational technologies cannot fully unlock student learning potential by tackling all diversity facets during different learning activities. They cannot indicate how specific learning outcomes result for a student with a certain background, goals, motivation, and reasoning habits. In general, there is little insight on the diagnostics (causality) between student learning and teaching.  Also, current algorithms and technologies cannot exploit the diversity of a team to educate students about important present issues, like peer assistance and mentoring, or social responsibility. 

Similar to a gym app that guides through different training regimes and intensity levels, we propose to devise a novel computer application that uses data analytics and ML to measure and causally explain (dynamic diagnose) the progress and shortcomings of student learning, and then utilizes the insight gained on individuals to optimize team activities, like course projects. The proposed environment focuses on assessing student learning and diagnosing their difficulties and misunderstandings about the taught concepts. Diagnosis pertains to dynamic diagnostic formative assessment, which aims to uncover the causes of learning shortcomings~\cite{Treagust, Bell2001, Bell2001a, Libarkin2001, Niss2012}. Dynamic diagnostic formative assessment belongs to constructivist theories in education~\cite{Adams1995, Bell1996}. These theories consider students are active participants in shaping the way in which knowledge is presented and learned. The proposed system groups learning difficulties into four categories based on insight on learning proposed by studies in cognitive psychology~\cite{Keene2010, Britto_Usman_2015, Piaget2008, Sharunova_Butt_Qureshi_2018}. These categories are recall from memory, concept adjustment, concept modification, and problem decomposition into sub-goals (sub-problems) and concept combination.  Data models are proposed to predict the occurrence of each of the four challenge types, as well as their connection into an overall model that represents a student's learning trajectory. In addition to cognitive aspects, the paper argues that emotional and social behavior during learning must be included too. Then, the models can be used to automatically create real-time, student-specific interventions (e.g., learning cues or even additional exercises and reading materials) to address less understood concepts and topics.

The goal of this work is to create a novel system based on a cognitive, social, and emotional Machine Learning (ML) algorithms to monitor students' cognitive and affective states and to personalize their learning experiences in and outside the classroom. We envision that the system will enable the creation of new adaptive pedagogical approaches to unleash undergraduate computer engineering and science student learning potential through two main objectives: customize the course material based on background, abilities, situation, and progress of each student; and leveraging diversity-enabled learning experiences.

The paper has the following structure. Section~2 discusses related work. Section~3 presents the limitations of traditional assessment and proposes diagnostic assessment as a way to pinpoint the specific learning difficulties of a student. Section~4 discusses the methodology of the automated system for dynamic diagnostic and intervention. A discussion of the system based on examples follows next. Conclusions end the paper.

\section {Related Work}

Learning is a complex process that depends on a student's state over multiple dimensions as well as on the state's dynamics over time. The student's cognitive (e.g. background knowledge), metacognitive (e.g., motivation and goals), and affective  (e.g., joy, engagement, boredom, frustration, anger, confusion) state before and during learning affect learning outcomes \cite{Worthal:2019, Sinclair:2018, Taub:2019}. Being able to monitor in parallel the states along these dimensions during a learning activity offers rich insight into how engaged a student is, how frustrated, confused, bored, irritated  in relationship to specific concepts or assignments. This information can be used in two ways: (1) to create student learning profiles, i.e. what they know, what they have difficulties with, how do they react when they have problems, and (2) to create an aggregated class profile that clusters students into different learning categories~\cite{Anand:2018}. To close the loop and positively affect learning, both the individual and class profiles can be fed into different adaptive feedback mechanisms, at the individual and team level - propose hints, ask for hints from more advanced students, propose alternate exercises or study materials, at the class level, inform teacher of which concepts and assignments have created the most problems, help identify misconceptions in concept understanding and use. The class profile can be used to identify best study groups for students from different clusters of learning for each assignment.

The grounding theory of the proposed methodology for automated diagnostic assessment and intervention creation is Bloom's Taxonomy~\cite {Keene2010, Krathwohl_2002, Britto_Usman_2015, Sharunova_Butt_Qureshi_2018}, and its deep connections to the large body of work originating from Jean Piaget's theory of learning \cite {Case1985, Piaget2008} and to Gardner's theory of multiple intelligences \cite {Gardner2012}. The large body of work originated from Piaget's theory of learning considers that human learning follows a sequence of successive cognitive stages: (1) sensorimotor stage (when new concepts are learned through their physical, material features), (2) symbolic stage (when symbols replace the physical facets of the learned concepts), (3) intuitive thought stage (when incipient reasoning starts to utilize the new concepts), (4) concrete operational stage (when logical reasoning uses the learned concepts), and (5) formal operational stage (when abstract reasoning, including creative problem solving, employs the new knowledge to create new solutions). The five stages can be also linked to the seven principles of Ambrose et al. \cite{Ambrose2010}: prior knowledge, knowledge organization, motivation, mastery, practice and feedback, course climate, and self-directed learning. There are also intriguing connections to other learning theories used in modern educational technology development, like the seven principles of Ambrose et al. \cite{Ambrose2010}. 

Bloom's Taxonomy has been widely used to propose new pedagogical approaches to engineering teaching \cite {ABET}. It is also the fundament for ABET's Engineering and Computer Science student outcome assessment procedure. ABET is the main organization involved in program accreditation. Bloom's Taxonomy defines six cognitive levels: (1) knowledge (includes recognizing newly learned concepts), (2) comprehension (involves demonstrating the understanding of learned concepts), (3) application (implies using the acquired concepts in new situations), (4) analysis (is the decomposing a concept into its pieces and understanding the relations between pieces), (5) synthesis (represents creating a new pattern using the learned concept), and (6) evaluation (is the critical judging of concepts). 

Many studies have estimated the affective states of students during learning activities such as computer-based tutoring, individual, collaborative or classroom learning settings. Methods range from employing a multitude of sensors (camera for face recognition, gaze tracking, skin conductance, EEG, temperature, heart rate, body posture and movement, etc.) \cite{Lane:2019, Dmello:2018, Bosch:2016}, to verbalizing emotions \cite{Craig:2008}, to surveys \cite{Worthal:2019, Sinclair:2018, Taub:2019}. Sharma et al. 2019 \cite{Sharma:2019} has tracked the gaze and the face, and monitored body physiological response using heart rate, electrodermal activity, temperature, blood volume pulse sensors, EEG while students were taking an adaptive computer test. They used feature selection methods to find the set of features most predictive of effort and assessment performance. Among the most informative features were number of eye fixations, average saccade duration, intensity of nose wrinkles, and from the wristband sensors, average temperature. This approach aggregated the sensors data over time and did not synchronize the affective and cognitive states with the student's task. It was observed that the accuracy of affective state estimation improves by using multiple modalities (e.g. face recognition and eye tracking) over a single sensor channel \cite{Worthal:2019}. Also, physiological measurements of emotions alone lack ground truth - i.e. they assume a direct relationship to feelings of engagement, confusion, boredom. But previous studies show weak correlations from sensor measurements to feelings \cite{Sharma:2019, Dmello:2018}.
 
Several approaches tracked the dynamics of affective states over time \cite{Sinclair:2018, Dmello:2012a} using periodic surveys or eye gaze sensing \cite{Jaques:2014} during a student's interaction with a computer tutoring system. It was observed that both positive and negative feelings can be beneficial to learning \cite{Worthal:2019, Sinclair:2018}. Positive effects of confusion as a measure of cognitive dissonance, that once resolved can lead to a deep learning state \cite{Sharma:2019, Craig:2008}. Approaches such as Hidden Markov Modeling \cite{Afzal:2006}, Gaussian Process Classification \cite{Kapoor:2007}, Bayesian network models \cite{Sam:2012}, Dynamic Decision Network \cite{Conati:2009}, or Latent transition analysis (LTA) \cite{Sinclair:2018} have been used to identify affective states and transitions from a state to another. For example, in \cite{Sinclair:2018} they identified three states: positive, boredom/frustration and moderate and found common transitions from moderate to boredom/frustration, and not common transitions out of a boredom state. Craig et al. 2004 \cite{Craig:2004} showed that students who were not able to transition out of a frustration state had a less effective learning outcome.  Ability to identify the state and its transitions is an essential ingredient to an intelligent system which can then adaptively inject a hint or an alternate path on or before a transition to a negative state takes place \cite{Sinclair:2018}. These approaches suggest there is information in state transition, but so far this has not been connected to the learner's cognitive state (e.g. what is the activity that caused a state transition) and further to an active feedback mechanism. 
 

The proposed methodology represents a Cognitive - Emotional - Social (CES) Machine Learning (ML) environment that characterizes (models) student learning according to the five stages of Piaget's theory, and then causally relates (diagnostics) the models to the observed teaching outcomes pertaining to the six levels of Bloom's Taxonomy. Hence, from a theory of learning point of view, the methodology aims to explain the cause - effect relation between student learning in a course and his / her acquired knowledge. It also plans to capture the connection between the emotional and social student behavior during team activities (e.g., according to theories on emotion modeling and team behavior) and the cognitive theories on learning (i.e. Piaget's theory and Bloom's Taxonomy).      
            
The methodology is conceptually similar to educational systems based on theories in cognitive psychology. For example, Cognitive Tutor \cite{Ritter2007, Vincent2002} relies on the ACT-R cognitive architecture \cite{Ritter2019}. ACT-R emphasizes memory modeling and knowledge representation as associative and procedural components. The discussed CES-ML methodology is based on our recent work on the InnovA cognitive architecture for design innovation \cite{Li2018}. InnovA's contribution is its focus on open-ended problem solving through causality mining and concept combination. This corresponds to the concrete operational and formal operational stages in Piaget's theory and to analysis, synthesis, and evaluation levels in Bloom's Taxonomy. These upper stages and levels are less emphasized by other cognitive architectures. InnovA also considers emotional and social interactions, which are important for team activities, but were not considered in previous work on cognitive-based educational technologies. The novelty of the proposed methodology as compared to other ML-based educational technologies relates to the upper learning stages, logical thinking and formal operational stages in Piaget's theory and analysis, synthesis, and evaluation in Bloom's Taxonomy. The early stages of the methodology follow the approach also used by current technologies, i.e. building regression or Deep Learning models for student learning.

\section {Student Learning}

\subsection {Limitations of traditional assessment}

We referred in this paper to ``Programming Fundamentals'' course (ESE 124) at Stony Brook University, even though our discussion is more general. ESE 124 is a required freshman course on C~programming. It is attended each semester by about 120 students. It teaches both basic and related concepts, like iterative algorithms, bit-level operations, arrays and structured data types, file operations, abstract data types, pointers, and dynamic memory management. Scheduled weekly lab activities complement the lectures.  

We have assessed student learning over three consecutive years. Traditional methods were used, including those recommended by ABET, the organization granting engineering programs accreditation~\cite{ABET}. Used methods included a mixture of grades assigned to quizzes, homework, laboratory assignments, course projects, and exam questions. Grades and justifications for the grades, like the mistakes made by students, are tools through which instructors can assess student learning progress and the degree to which course objectives and expected student outcomes are attained. Moreover, ABET's continuous improvement criterion suggests using grades, called direct assessment methods, as the main tool to identify systematic shortcomings in a course or program, and then to accordingly address them to improve course content or program. 

In addition to standard assessment metrics, like average grade and percentage of students meeting an outcome threshold set for the course, a variety of other data analytics were computed based on the received grades: averages for the student and class, minimum and maximum grades, multiple quartiles, like first (Q1), second (Q2), and third (Q3) quartiles, and outliers. We also computed metrics that capture trends for individual students and the entire class and produced comparisons with peers or other previous student groups, e.g., student from different majors or different years. Finally, students were clustered based on these metrics, like their grades, or based on their learning evolution (trend) over time. Moreover, student groups were clustered over time to discover systematic trends. While employing data analytics is still not ubiquitous, it has been intensely studied for assessing teaching and student learning~\cite{Huang2013, MacFadyen2010, Oliva2021, Sergis2017, Xing2015}. 

Still, data analytics and traditional assessment metrics and methods offer only weak diagnostic insight into the causes that resulted in poor outcomes and ineffective learning, e.g., learning misconceptions~\cite{Bell2001, Bell2001a, Treagust}. It has been impossible to precisely and systematically find if cause were the specific methods of knowledge delivery during teaching, or the nature and difficulty of exercises, homework or lab activities, or insufficient student background, or motivation, or a mixture of all the above. For example, computing weighted averages of grades (a popular ABET direct assessment method) often indicates that course objectives were met. However, averages often mask low grades or systematic learning inefficiencies. For examples, the populations of students in two different majors but taking ESE 124 at the same time had very similar averages, but one had a bimodal distribution of grades, while the other had unimodal distribution. Student background was not correlated to this difference. Metrics, like quartiles, can expose other imbalances in student populations, like splitting of the population into clusters with low, average, or high grades, but the causes of these observations remain unclear. Other observations remained equally unexplained, like the relative high number of students with many uncompleted homework, or the main difficulties in completing more complex programming exercises, including the course project.  

The precise reasons that produce these imbalances usually remain unknown. While ad-hoc explanations, like insufficient student background or too fast lecturing pace or too difficult exercises, might obviously be causes, correlation analysis is often weak and non-uniform. Other reasons, like improper workload balancing, high anxiety, and loss of motivation are important too. Moreover, other results, like grades being skewed towards low or high grades, the number of low and high outliers (i.e. maximum scores), or the wide dispersion ranges of grades are hard to systematically explain based on traditional metrics without insight about the entire learning process. Different learning dynamics can produce similar cumulative values in traditional assessment based on grades. Direct assessment can give only a partial image. For example, exam questions are often insufficiently robust to reflect the degree to which students attained the learned concepts, as during exams they must effectively balance the exam time towards solving three or four questions. Homework and lab assignments might not fully reflect a student's individual level of knowledge attainment. Some students seem to be happy with meeting the minimum requirements of the course.

In summary, teaching and learning assessment must be extended beyond the capabilities of traditional methods to offer robust and early diagnostics into the reasons of learning shortcomings. Systematic diagnostics can highlight the needed, student-oriented changes, adjustments, and additions to improve learning. Assessment should be flexible to address a wide range of goals, e.g., minimizing the amount of learning loopholes and misconceptions, meeting goals set for the population, emphasizing error-free problem solving, or minimizing the number of students that target only minimum course requirements. This vision requires a fundamental change in the assessment approach from assessing only the outcomes of learning, e.g., graded course work, to assessing the entire learning process that produced the course work. Also, assessment must observe the variety of dimensions of student learning, such as cognitive, social, and emotional. Often a student's progress depends on social and emotional aspects, like motivation and an understanding of how the individual course work ties to the professional career. Student motivation decides the willingness to spend time and effort on learning. Moreover, assessment should uncover systematic trends and causes at the level of a student population. Used methods should be scalable for large, diverse populations while still being able to identify individual student needs. 

\begin{figure}[t]
\centering
\includegraphics[width=6.7in]{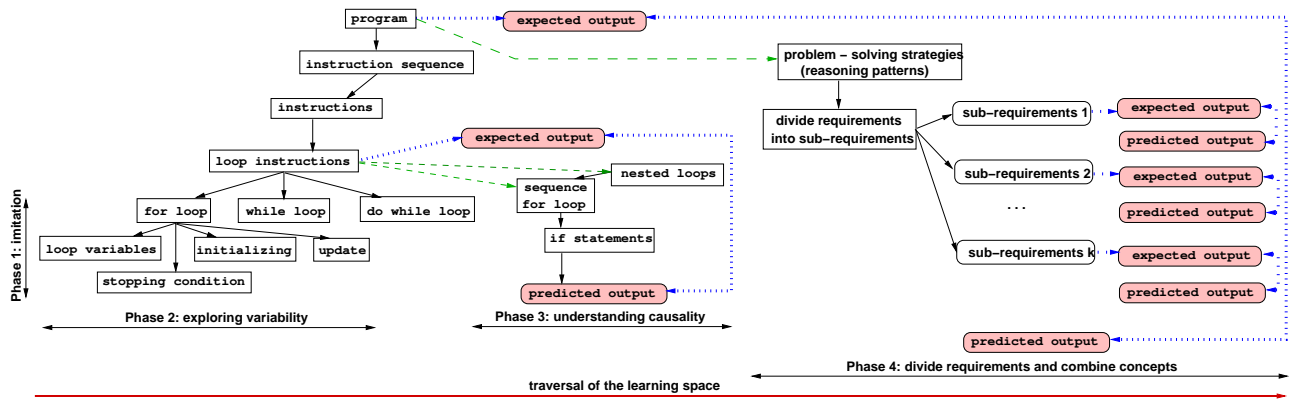}
\caption{Learning space fragment for ``Programming Fundamentals'' course}
\label{learning_space}
\vspace * {-0.1in}
\end{figure}

\subsection {Proposed dynamic diagnostic formative assessment}

Constructivism considers that learners actively create their own new knowledge structures based on their own reality and experiences, including background and previous events~\cite{Adams1995, Bell1996}. Constructivist theories assume that students build sensible and coherent understandings seen through their own experiences and background, even though this understanding might not match the intended meaning taught by an instructor~\cite{Treagust}. Hence, the goal of constructivist approaches in education is that teachers and students create a collaboration environment, in which students are actively engaged in their own learning.  Instead of being just passive receivers of concepts taught in a course, they participate to the specific shaping of the way in which new knowledge is presented and learned. A central component of constructivist learning is diagnostic assessment, as it not only measures students' level of attainment of the taught material, but also points towards the causes for learning shortcomings. This insight is then used to accordingly modify and improve the course material.   

In diagnostic formative assessment, finding the roots of knowledge (concepts) misunderstanding requires a sequences of (i)~interventions posed to a student, e.g., asked questions,  followed by (ii)~the analysis of the offered answers~\cite{Bell2001, Bell2001a, Libarkin2001, Niss2012, Treagust}. Existing diagnostic assessment methods use surveys, like two-tier surveys, which include all expected answers that students might offer (both correct and incorrect), as well as their explanations for their answers. Incorrect responses might point towards errors in understanding (explicit causes). However, causes, like insufficient background, might not be directly reasoned from responses (implicit causes). Addressing both explicit and implicit causes require a sequence of interventions (i.e. asked questions and response analysis) with the goal to converge towards the true cause for a student's incorrect understanding. This process needs synthesizing successive hypotheses about the cause, and then validating or invalidating them based on the student's responses. It is possible that the process does not converge, hence, the nature of incorrect learning is not found, at least momentarily.   
 
Gaining insight about the causes of learning shortcomings allows then to redesign and adjust the curriculum, so that learning shortcomings are mitigated. Well calibrated, targeted interventions by instructors, peers, or self-correction can help a student converge towards learning the taught knowledge as intended by the instructor. 

The nature of learning misconceptions and hence diagnostic assessment depends on the specific cognitive phases of student learning. A summary of the successive cognitive learning phases~\cite{Piaget2008, Case1985} for ``Programming Fundamentals'' course (ESE 124) was shown in Figure~\ref{learning_space}. The first phase is {\em recall phase}. Students imitate exercises presented in class or in textbooks with the goal to correctly recall the taught concepts. They learn the ``physical facets of new concepts'', i.e. syntax and experience the execution (behavior) of the concepts for different parameter values. The second phase is {\em adjustment}, in which student explore the variability of a new concept. Like during the symbolic stage \cite {Piaget2008}, students consider different variations of a new concept (i.e. different conditions to express loop continuation, loop nesting) to produce their own cognitive image for the encompassing (abstract) concept. The third phase is {\em concept modification}. Similar to intuitive thought and logical thinking stages \cite{Piaget2008, Case1985}, students focus on understanding causality, e.g., how a concept creates its execution trace and output by being connected to the other concepts of a solution. Finally, phase four is {\em concept combination}. Students target solving problems by dividing requirements into simple activities, and then combining the related concepts. This step is like the formal operational stage~\cite {Piaget2008}.  

In our assessment of student work for ESE 124, we observed that their learning shortcomings pertained to the four phases:
\begin{enumerate}
\vspace * {-0.15in}
\item
{\em Incorrect recall}: Students incorrectly recalled programming concepts taught in class, like syntax and semantics (meaning) of fundamental C constructs.

\vspace * {-0.05in}
\item
{\em Incorrect adjustment}:  Students correctly found the concepts necessary to solve a problem. However, they incorrectly identified the needed adjustments for the new problem, and / or erroneous predicted the effects of the changes. 

\vspace * {-0.05in}
\item
{\em Incorrect modification}: While students could relate the new exercise to problems presented in class, their understanding of the causal (logic) connection of the programming constructs into a solution was incorrect, or they could not go beyond incremental changes of the taught exercises.   

\item
\vspace * {-0.05in}
{\em Incorrect decomposition and concept combinations}: Students could not relate the new exercise to the exercises or concepts discussed in class. Hence, they were not able to start solving it, saying that they ``feel stuck''. While students learned the individual concepts, they were unable to identify or causally relate them to create solutions for a new problem.
\vspace * {-0.1in}
\end{enumerate}  

\begin{figure}[t]
\centering
\includegraphics[width=6.7in]{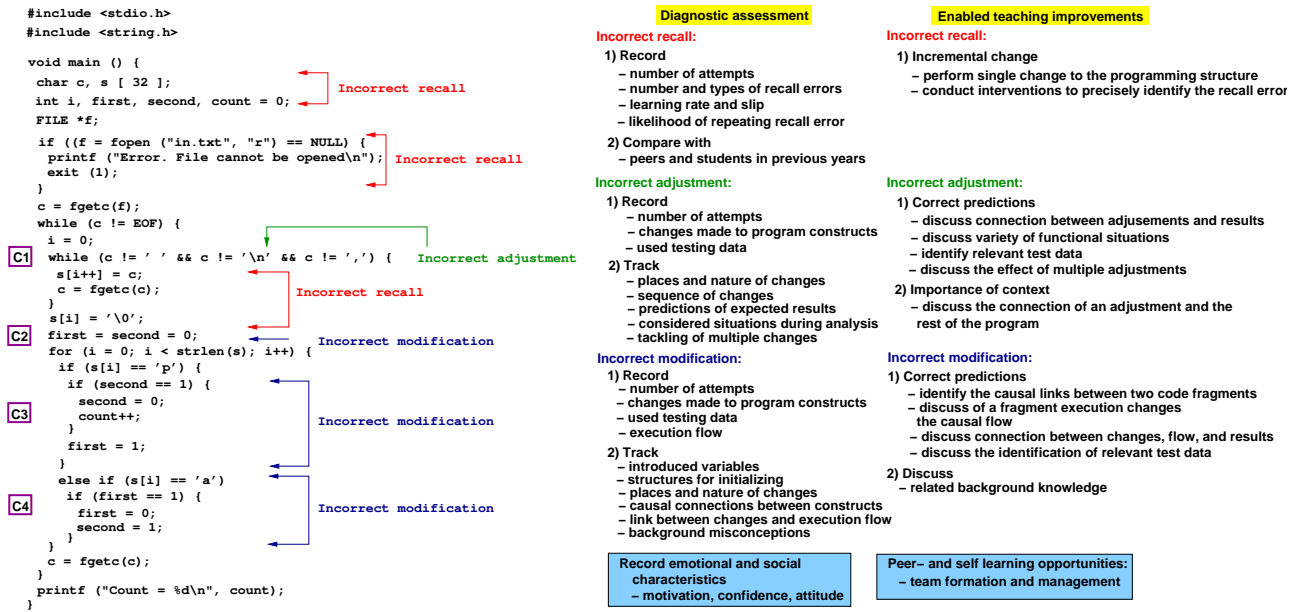}
\caption{Sample of student learning shortcomings, related diagnostic assessment, and enabled teaching improvements}
\label{sample}
\vspace * {-0.1in}
\end{figure}

\subsection {Case Study} 

We detailed next diagnostic assessment for the four types of learning shortcomings. We referred to a homework exercise that asked students to count the number of occurrences of pattern "pap" in the words of a text file. Overlapping patterns were allowed. Figure~\ref{sample} shows the reference code, related learning misconceptions, diagnostic assessment, and enabled teaching improvements. 

\underline{\em 1. Incorrect recall}. The first learning phase reinforces the correct recall of newly learned knowledge~\cite{Piaget2008}. In Figure~\ref{sample}, recall (shown in red) refers to variable declaration, opening a text file in the read mode, and reading a word from the input file and storing it in a string. These concepts were discussed in class and students were required to correctly recall them.   

{\bf Diagnostic assessment}. Assessment measures a student's capability to correctly recall the concepts presented in class and counts the number and types of recall errors. Other metrics include the number of attempts until fixing all recall errors in an exercise, learning rate and slip, and the predicted likelihood of repeating the recall errors in the future. 

The measured parameters and all predictions become part of a student's learning aptitude (SLA) model. SLAs serve to compare a student's performance with that of his / her peers or from previous classes. 
 
{\bf Enabled teaching improvements}. Targeted interventions, e.g., similar exercises, more explanations, tutoring, suggested readings, etc., can be devised for each student. For example, similar exercises van be devised by incrementally changing taught exercise by preserving their structure, but changing the parameters of the structure, e.g., the variables and operators in expressions and conditions, or the logic operators used in defining the conditions of $if$ and $loop$ statements. Each similar exercise has a single change, as the goal is to improve correct recall of newly learned concepts.
 
\underline{\em 2. Incorrect adjustment} occurred when students correctly identified the concepts to be changed and the structure that uses them (e.g., the $while$ loop in green in Figure~\ref{sample}), but the instancing of the concepts and / or the specific connection to the rest of the program had errors. For example, students incorrectly selected the possible separators between the words in a text file. Errors were due to misconceptions about particular concept features (e.g., whitespace ASCII characters), and incorrect predictions of the adjustment results (i.e. handling of separator characters). Some errors were because of the dependencies between multiple adjustments. The observed difficulties suggest an insufficiently detailed cognitive image about newly learned concepts to reflect their possible variations and corresponding effects.    


{\bf Diagnostic assessment}. Assessment characterizes a student's ability to correctly identify the required concept changes and their implications, including their sequenced tackling from simpler to more complex changes, if multiple adjustments are needed. Addressing simple changes first is motivated by that it can help a student's understanding of the concept and is likely to improve his / her confidence and motivation. Interventions, i.e. posed questions, are produced for each adjustment, its connection to the surrounding code, related test data, expected results, and links between multiple changes. Data acquired during diagnostic assessment includes the variety of tackled concepts and proposed adjustments, changing the code surrounding the adjusted concepts, identified test data, predicted effects of changes, and adjustment errors.  

Assessment creates description models of the degree to which a student is likely to suggest concept adjustments for similar or distinct exercises. Predictions on the expected learning rates are also possible. Models are updated every time the same concept has been used in new solutions, hence giving an image of the learning progress over time. Moreover, the description presents situations that were not part of the used test data, and thus can possibly be cases that the student might find it hard to solve in the future. The acquired data and the models become part of a student's SLA.

Errors between expectations and correct results highlight the situations of adjustment misunderstanding. 
{\small
\begin{equation}
Errors_{concept~i} = \sum_{\delta_{i,j}, \Delta_{i,j}} || Result (concept_i + \delta_{i,j}) - Correct~result (concept_i + \Delta_{i,j}) ||^2
\label{eq1}
\end{equation}
}
where $\Delta_{i,j}$ are the expected adjustments~$j$ of concept $i$, and $\delta_{i,j}$ is the performed adjustment. The challenge for diagnostic assessment is the statistical selection of the representative samples~$\Delta_{i,j}$, especially considering the broad variety of contexts in which the concept can be used in a solution. It is not realistic to assume that all contexts can be covered during selection. Instead, contexts must be prioritized based on their frequency of occurrence for the course and related courses. 

{\bf Enabled teaching improvements}. Using diagnostic assessment data, teaching can focus on explaining the associations between concept adjustments and the produced output changes. Another option is to use diagnostic data to teach students to reason through analogy with previously taught examples and exercises. The purpose is to reinforce how to extend a taught reasoning sequence to address a similar exercise, hence, to create a mental image about the flexibility of a concept in accommodating new requirements. This process is similar to analogical reasoning~\cite{Balazs2002, Daugherty2008} as well as to problem solving through backward reasoning~\cite{Barbero2020}. Teaching can also address certain learning habits and improve motivation. 

Teaching must stress the importance of exploring the variety of the functional situations in which a new program solution must operate, the related concept adjustments, and the effect of adjustments. For example, students must understand how incremental changes to concepts produce new results, like how new $if$ conditions, $loop$ stopping criteria, or array indexes relate to iterations and final results. They will build more complete mental images of a new concept meanings by being able to correctly predict the effect of concept adjustments. This learning goal is achieved by selecting representative test data and explaining the differences between the changes expected by a student and the correct results. The dependency between concept adjustments and the rest of the program can be also detailed starting from diagnostic insight. Another goal is to teach on finding a broader set of adjustments to solve a problem, not just one solution. The pros and cons of each alternative are discussed and compared to connect the mental images of the alternatives. While the analysis is optional here, it can be beneficial for more complex learning steps, like to create new or modify previous algorithms.  


\underline{\em 3. Incorrect modification} refers to situations in which misconceptions relate to students' incorrect understanding of the causal (logic) connections of the programming constructs into a solution, thus students cannot reason beyond simple, incremental changes of the taught concepts and exercises. 
Specifically, students cannot take a previously taught exercise as a starting point, and change its execution flow and processing, so that new algorithmic requirements are correctly met. For example, for the sequences shown in blue in Figure~\ref{sample}, students had difficulties in understanding how the execution flow of the instructions must be controlled, so that all instances of the searched pattern were counted. They made errors in identifying what variables are needed to describe the control, where to place them in the program code, how to iteratively modify them, and what data processing to associate to each flow modification. A few instances were observed in which students had incorrect understanding of character representation as ASCII code, a topic discussed in a pre-requisite course. Students were initially convinced that their understanding of ASCII representation was correct, and hence strictly focused on the newly taught concepts. However, they were able to uncover the root of their difficulties only after spending significant time on watching unstructured information on the Internet to finally find the root of their learning difficulties. 

Assessing incorrect modifications is hard, as it involves reasoning with causal (logic) connections of multiple concepts in diverse situations. We observed that one-on-one sessions with the instructors do not scale well for larger classes, because they require a long time. Initial discussions were often only marginally related to the concrete learning misunderstandings, being masked by insufficient background, e.g., ASCII representation. Note that this deficiency was not uncovered through traditional assessment but was anecdotally revealed through short student Op-eds to describe difficult learning situations.    

{\bf Diagnostic assessment}. For the exercise in Figure~\ref{sample}, student learning during concept modification can be expressed by the following relation, which expresses the sequence of related learning steps:
{\small
\begin{multline}
Understanding \sim [ Identify~all~situations + \Delta_1] \prec [ Identify~variables + \Delta_2 ] \prec [ Initialize~variables + \Delta_3 ] \prec \\ [ Find~places~to~modify + \Delta_4] \prec [Modify~variables + \Delta_5] | [ Background~information \leftarrow <\Delta_1, \Delta_2, \Delta_3, \Delta_4, \Delta_5> ] 
\label{eq2}
\end{multline}
}
Operator~$\prec$ indicates precedence in learning a concept, operator~$|$ defines the context (e.g., background) in which the problem is solved, and operator~$\leftarrow$ represents the instancing of the background for the specific concept modifications $\Delta_1$, $\Delta_2$, $\Delta_3$, $\Delta_4$ and $\Delta_5$. The relation expresses the cognitive dependencies and adaptations needed to solve the problem. 

As shown in relation~(\ref{eq2}), assessment must tackle two aspects: (i)~addressing incorrect understandings about the meaning of related concepts and their incorrect connection into solutions (expressed through operators $\prec$, $|$, and $\leftarrow$), as well as (ii)~the specific changes needed for concepts and the overall structure. Specifically, for the exercise in Figure~\ref{sample}, equation~(\ref{eq2}) states that diagnostic assessment must address five successive aspects: (1)~It should assess the degree to which a student identifies all situations in which the searched pattern occurs in a word, like at the beginning, inside, and the end of the word, or several successive pattern overlapping. (2)~It characterizes the degree to which a student understands the required variables to correctly tackle the found situations. (3)~The variables must be initialized. (4)~Assessment must describe a student's capability to locate the places in the code, where modifications are needed in order to solve the new problem for the identified situations. (5)~The correctness of the proposed variable modifications must be assessed, e.g., the execution flow of the program. The five specific modifications are expressed as $\Delta_i$ in equation~(\ref{eq7}). Finally, the assessment must be conducted in the context of a student background acquired in the pre-requisite courses, like the meaning of ASCII code. 

Data collected during diagnostic assessment refers to the variety and correctness of the identified causal (logic) relations between code fragments, reasoning about the expected execution flow, and completeness of used test data. Data is then used to create models that predict the likelihood of correct causal identification and execution flow for similar and different exercises, and the more likely reasoning errors. The collected data and the produced models are added to a student's SLA.  

{\bf Enabled teaching improvements}. Diagnostic insight about the above five activities support student-targeted changes of the teaching process and material. The insight must be used to target precise steps of deductive reasoning for tying together code fragments, as well generalizations to understand the overall picture of the modification. Student-specific input must teach how the features and concepts of different exercises discussed in class must be combined or linked together to solve a new problem. Comprehensive mental images must be formed through exercises and peer-tutoring in teams that focus on the causal relations between code modifications, the resulting execution flow, and program results for test data. It includes detailing the causal links between related code fragments that were changed. Also, the changing of the execution flow for alternative code modifications must be highlighted. Discussion of possible flow changes must be connected to finding the test data to trigger those flows. The review of related background knowledge must be also added. 

Learning how to predict complex modifications is important especially for exercises that require multiple concept modifications. Complexity can vary within broad limits. For example, the pattern recognition exercise might initially require finding only nonoverlapping patterns. The control structure shown in blue in Figure~\ref{sample} is then simpler. Besides requiring overlapping patterns, a more complex exercise requests finding multiple patterns that are inputs to the program. 

\underline{\em 4. Incorrect decomposition and concept combinations} refers to situations in which students did not know how to start solving an exercise. Typically, they would indicate that they felt stuck, hence had difficulties even asking questions. Still, in the situations in which students could finally solve the problem, they mentioned that a preliminary step was identifying analogies or metaphors that helped them understand the problem by giving them a familiar image, e.g., a three-tier pocket to describe the abstract data type stack, or a building structure as a metaphor for structured data type, or a water pipe as an analogy for a queue. The mentioned analogies and metaphors might be student specific, meaning that other students might not necessarily find them useful in learning. Still, being able to connect the new knowledge to a student's prior framework is known to be a necessary learning step. Others relied on grit and determination to consult several textbooks, websites, or video archives to comprehend concepts. 

An unexpected insight offered by assessment through Op-eds on difficult homework exercises described student emotions during learning, i.e. their anxieties and frustrations. It is interesting to point out that anxieties were often preconditioned by previous experiences, like difficulties with Taylor series expansion in calculus, or comments made by seniors or relatives about topics, like pointers, arrays, or abstract data types. Left unaddressed, such prior beliefs and emotions can lead to dramatic results, like a student deciding to change his / her major. Two students mentioned in their feedback that they contemplated this option, even though their final performance was good.

{\bf Diagnostic assessment}. Assessment targets the degree to which students can combine previously and newly learned concepts to solve new types of exercises that are beyond the types of exercises solved in class. Assessment mines and characterizes the (i)~causality of the pattern elements and (ii)~the nature of pattern combinations. Besides reasoning, this process includes a step of trial-and-error for (iii) pattern adjustment to solve the new problem. 

The recorded data includes the effort, progress, and correctness learning as described by activities (i)-(iii). They capture the capacity to break down a problem into sub-problems, the errors during problem decomposition, the type of identified and used patterns, the kind and complexity of the combined patterns, the amount of trial-and-error to figure out details, and the amount of incremental variability used in adjusting the general patterns. The recorded data captures the student preferred ways of exercise solving, including decision making. Prediction models stored in SLA describe a student's cognitive ability to effectively decompose a new problem into sub-problems and produce solutions by combining and adjusting taught exercises. A parameter of the models is the degree of similarity of the new and the presented exercises. 

{\bf Enabled teaching improvements}. Research shows that solving complex problems requires (i)~breaking down the problem requirements into sub-goals / activities, (ii)~finding algorithmic templates for each sub-goal, (iii)~adjusting the templates for the new needs, and (iv)~combining them into a solution. Insight from diagnostic assessment should guide the process of teaching how to create the algorithmic structure (frame) for a new problem by decomposing the problem into parts and then composing the partial solutions (for the parts) into the overall solution. Teaching must help students create mental maps that aid in recognizing the processing frame and its parts to solve a new problem.

\subsection {Innovations of the proposed work in diagnostic formative assessment}

Our work proposes two main innovations to diagnostic formative assessment: dynamic and automated assessment. 

{\bf Dynamic diagnostic assessment}. The first innovation refers to the process utilized to identify the possible causes of learning misconceptions. Traditional approaches, like two-tier surveys, use a static dictionary and propositional statements to cover possible knowledge statements on a topic~\cite{Bell2001, Bell2001a, Treagust}. During assessment, students select from the static list of possible answers (multiple choice surveys), but in addition offer their own explanations for their choices. Possible answers include both correct answers and misconceptions. The gathered information helps building concept maps for a student to describe his / her level of understanding. While the static traditional approach has been successfully used in teaching high school physics, biology, and chemistry, setting up the static database required by the method is long and difficult. Moreover, periodically taking the two-tier survey is disruptive. Instead of a static list of possible answers and required student explanations, we propose a method that creates predictions to express the most likely student explanations including their understanding of the involved concepts based on successive samples of their course work, e.g., homework and lab assignments. The method does not request a large static library of possible answers, because it assembles misconceptions as the concept level (e.g., programming constructs) to produce predictions about misconceptions at the solution level.        

{\bf Automated diagnostic assessment}. The second innovation is in that we propose to automate the process of diagnostic formative assessment. The diagnostic assessment process of posing questions to students and then analyzing their responses in attempt to find the cause of their misunderstanding is similar to a computational dialog system, in which two agents communicate until they reach a consensus, i.e. the cause of the learning misconception. Specifically, the two agents interact through questions and answers until they reach shared knowledge, while identifying the information needed to bridge the knowledge gap between the learning agent and the agent posing the questions. The shared knowledge describes the concepts that were correctly learned by a student, while the found bridges describe aspects that an instructor must subsequently clarify, as they are points in which a student's understanding deviates from the expected learning. 

Computationally, the process is similar to the generator and discriminator networks in Global Adversarial Networks (GANs)~\cite{Goodfellow2014}, i.e. the generator represents a student providing answers and the discriminator refers to an instructor posing entity in an attempt to diagnose the misconception. Depending on the nature of the gap, the mismatch between the two agents can be of the following kinds: (i)~partial agreement, if the misunderstandings do not change the main meaning of concepts, but there are differences with respect to specific features, details, and nuances, and (ii)~conflicts, if the two understandings oppose each other in main aspects. An additional outcome of the two-agent process is forming an image about a student's learning characteristics and performance, which then becomes part of the student's SLA description.




\begin{figure}[t]
\centering
\includegraphics[width=6.3in]{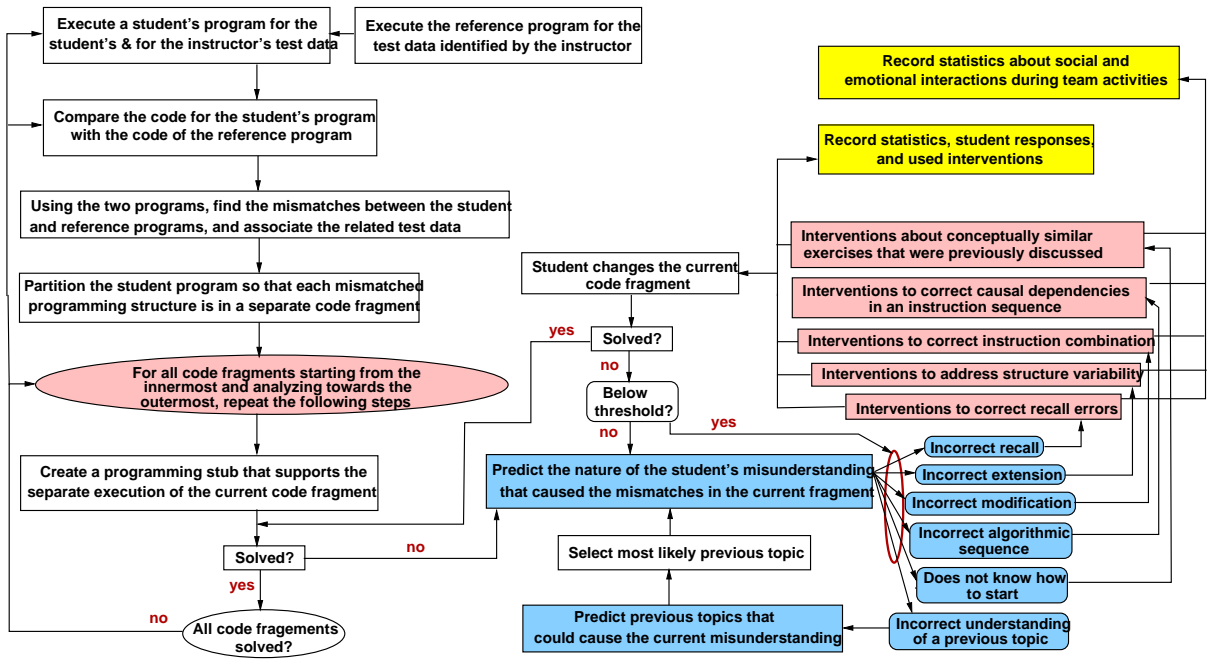}
\caption{Data acquisition process for diagnostic assessment}
\label{diagnostic}
\vspace * {-0.1in}
\end{figure}

\section {System Implementation}

This section discusses the overall flow for dynamic diagnostic assessment including the related data acquisition methods, the procedure to build Student Learning Aptitude (SLA) descriptions, and the automation of diagnostic assessment.   

\subsection {Dynamic diagnostic formative assessment for constructivist learning}

{\bf 1. Overall dynamic diagnostic assessment flow}. Figure~\ref{diagnostic} illustrates the proposed process for conducting dynamic diagnostic formative assessment without the need of setting up a costly, static dictionary and propositional statements, like for two-tier surveys. This is a major limitation of current diagnostic assessment. The initial steps in Figure~\ref{diagnostic} find the mismatches between a programming solution developed by a student and a reference program developed by the instructor. Mismatches are found by comparing the variable values throughout the code as well as finding the differences between the two programs. The found mismatches in the code and their associated variable values for the test data are then successively analyzed starting from the simplest towards the more complex cases, in the order of their execution. Each mismatched fragment is separated from the overall program, and incorporated into a stub, so that it can be executed separately. Depending on the nature of the mismatch, the process in Figure~\ref{diagnostic} predicts the cause of the mismatch among six categories: incorrect recall of a learned concept, incorrect extension of a concept, incorrect modification of a taught concepts beyond the situations presented in class, incorrect algorithmic sequence, difficulties on knowing how to start, and misconceptions on a previous topic (background). These are the main learning shortcomings discussed in Section 3.2. 

{\em Data acquisition for dynamic and automated diagnostic assessment}. The data collection during diagnostic assessment in Figure~\ref{diagnostic} includes four methods: monitoring the program code, gaze tracking, voice-based sensing of emotions and social interactions, and surveys and Op-eds. The four methods are summarized next. 

Monitoring the program code is achieved by periodical sampling of the code that is devised by a student or a team of students working together on the same program. The sampling period is a parameter of the environment, e.g., every two minutes or so. The changes performed on a program are automatically found by comparing two consecutive samples. 

\begin{figure}[t]
\centering
\includegraphics[width=6.7in]{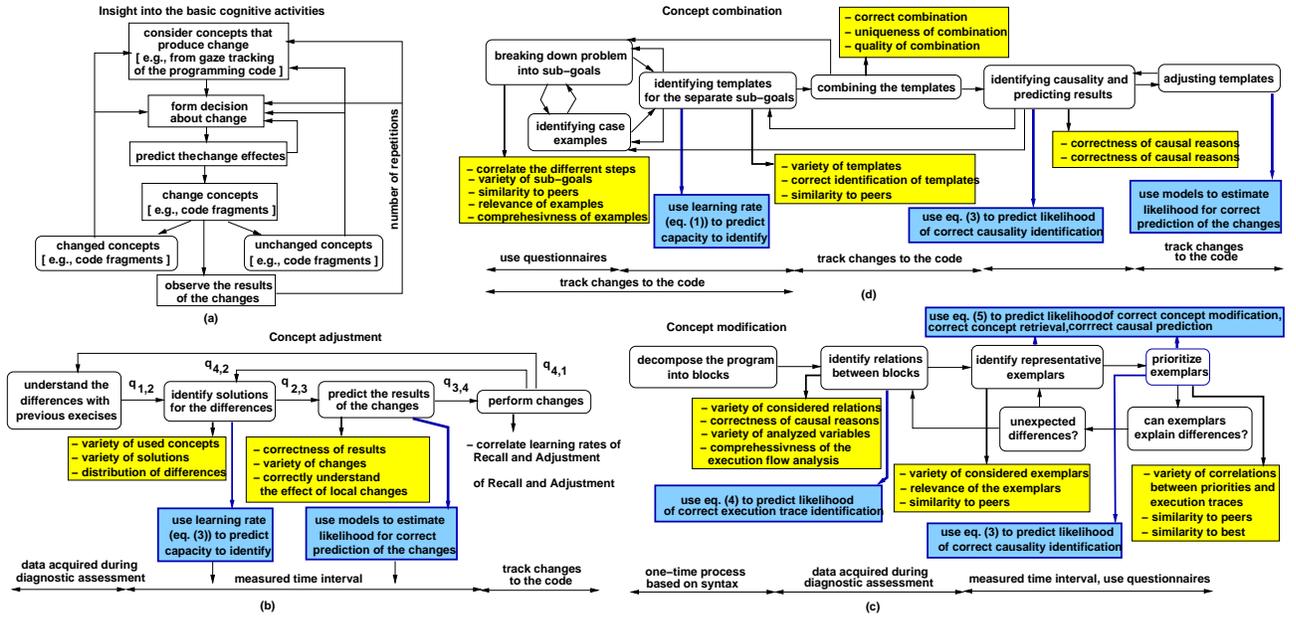}
\caption{Student Learning Aptitude description during diagnostic assessment}
\label{causality}
\vspace * {-0.1in}
\end{figure}

Gaze is continuously tracked, so that the environment records the code fragments that were analyzed by students during work. The position of the gaze is correlated with the current content of the foreground programming window, e.g., by using methods like {\tt GetForegroundWindow} to get a handle to the window of the Dev-C++ programming environment used in our ESE 124 course. Tracking the associations of gaze and corresponding code fragments over time enables to predict the nature of the expected basic analysis and reasoning steps performed by students before changing the code. The basic steps can then be related to the higher cognitive activities during problem-solving shown in Figure~\ref{causality}.         

The sensing of student emotions and social interactions between team members uses our previous work on automatically identifying and characterizing the emotions and interactions between group members using voice~\cite{Liu2015, Duke2021}. The designed system performs the following functions: (1)~speaker recognition, (2)~local data processing (e.g., tracking interactions between participants), (3)~emotion identification, and (4)~communication to the central server that collects the acquired data. The implementation uses deep neural networks to implement two supervised classifiers, one for speaker recognition and one for emotion identification. The system automatically identified the speaker, the amount of time he/she spoke, the sequence of interactions between participants, and the frequency of using keywords of interest. This information was used to characterize the interaction patterns between participants, the memes in discussions, and their sequence over time. The module was successfully used to recognize and track interactions involving more than three hundred participants working in teams of four students. The identified emotions include the following seven cases: Neutral, Anger, Boredom, Disgust, Fear, Happy, and Sad. Our experiments showed that the system has a very small error rate in identifying speakers, e.g., 96\% for speaker recognition and 87\% for emotion identification. 

Finally, surveys and Op-ed pieces are used to collect insight about broader beliefs and feelings, such as the degree of motivation for the course or the perceived utility of the course for the future profession. 
Surveys traditionally refer to specifics of homework and lab assignment exercises. Op-eds discuss thoughts about the learning difficulties encountered for a topic, approaches of tackling them, or various uncertainties and anxieties. 

{\bf 2. Student Learning Aptitude description during diagnostic assessment}. The data collected during dynamic diagnostic assessment is utilized to create Student Learning Aptitude (SLA) descriptions of each student. They summarize a student's learning activity, e.g., number and variety of attempted solutions, their correctness, number and nature of the made errors, the completeness of the used test data, and so on. SLAs also include a set of models predicting a student's learning and slip rates, likelihood to commit or repeat certain errors, likely success in solving certain types of problems, expected future difficulties in solving new problems, expected progress until the end of the course, etc.    

Figure~\ref{causality} summarizes the prediction model creation for SLAs. Figure~\ref{causality}(a) depicts the steps of the basic cognitive process that leads to a change of the code. The specific cognitive activities that correspond to the four different types of learning shortcomings and summarized in Figure~\ref{causality}(b)-(d) result depending on the nature of the performed analysis and reasoning. The first step identifies the concepts that are analyzed as potential elements of the changes that must be performed to the code. As a result, the students make a decision about the needed change and then performs the change. Finally, he / she observes the results of the change. The process can repeat a number of times.  

Gaze tracking traces the analysis sequence over time, e.g., the sequence of analyzed code fragments. This information supports setting up hypotheses about which cognitive activities were more likely to happen during the process of addressing specific learning misconceptions. The analyzed concepts, i.e. code fragments, are grouped into those that were changed and those that remained unchanged. The amount of time spent on each code fragment, the number of repetitions among the analyzed fragments correlated with the actual change suggest a likely reasoning flow that lead to the decision that produced the change. However, these parameters are different for various learning activities, like recall, adjustment, modification, and concept combination.

Note that this process goes beyond only understanding the correctness and faults of the logic reasoning process that creates a program. Insight is obtained about the cognitive steps that lead to an error in logic reasoning~\cite{Duke2021}, e.g., systematic insufficiency of certain cognitive activities that cause the observed errors. 

For small teams of three, four students work together on the same programming exercise, the process showed in Figure~\ref{causality}(a) must be mapped to the participating students starting from the observed social interactions and their emotional behavior. Students with an active participation have predicted cognitive activities close to the predictions based on the measured assessment data. The predictions for less active students are less depending on their degree of involvement.   

We discussed next the models for each learning shortcoming tackled during diagnostic assessment. 

\underline{\em 1. Incorrect recall}. The structure in Figure~\ref{causality}(a) can be instantiated as follows: Decision and prediction are less, most of the effort being on considering (recalling) concepts from memory and observing the results. Recalls and analyzing the results are mostly localized. 
Data acquisition during diagnostic assessment records the recalled concepts, the number and nature of recall errors (e.g., the dissimilarity with reference concepts), and the number of attempts to fix recall errors. 
Routines predict the correct retrieval of newly learned concepts (learning rate \cite{Pardos2010, Xu2017}) depending on parameters, like learning time, effort, number of attempts needed to successfully solve a problem. For example, the expected likelihood of a student to correctly use a concept is as follows:
{\small
\begin{equation}
Concept~recall () = M_{REC} (ConcS, ErrS, NrAtt, emotion, social)
\label{eq3}
\end{equation}
}
where parameter $ConcS$ is the set of concepts that must be recalled, $ErrS$ represents the number and types of recall errors, $NrAtt$ is the number of attempts to fix errors, $emotion$ describes the student's current emotions, and $social$ expresses his / her social interactions during learning. 

The likelihood of correctly understanding and recalling a concept is then used in equations (\ref{eq4}) and (\ref{eq3}) to estimate the likelihood of correctly solving more complex exercises. 

Current work uses off-the-shelf data analytics and ML algorithms, like ANOVA tests, t-test, correlation analysis, regression, deep learning, random forests, k-nearest neighbors, decision trees, and probabilistic graph models, to predict course grades, GPA, learning rate, learned state, at-risk students, and skill prediction\cite{Bekele2005, Hellas2018, Pardos2010, Xu2017}. For example, Hidden Markov Models are proposed to predict the acquired student skills based on learning rate, slip, and guess \cite{Pardos2010}. Bayesian belief networks are suggested to predict student performance depending on interest, confidence, motivation, attitude, and interest \cite {Bekele2005}. To complement existing work, the proposed models also incorporate emotional and social parameters in addition to the cognitive aspects for prediction of correct concept learning. They are known to be important too~\cite{Afzal:2006, Craig:2008, Sam:2012}.              

\underline{\em 2. Incorrect adjustment}. This step goes beyond correct recall of taught concepts, as students must incrementally change concepts and their using in an algorithm. Students must analyze the differences between the requirements of a new exercise and previously taught exercises, and then identify which concepts and adjustments of these concepts satisfy the new requirements. The structure in Figure~\ref{causality}(b) is mapped to Figure~\ref{causality}(a). As explained in Section~3 and illustrated in Figure~\ref{causality}(b), the step includes four activities that find the differences, identify possible adjustments, predict the adjustment results, and change the concepts. This step is performed iteratively until a correct solution is found, or the student stops as he / she cannot solve the problem. Referring to Figure~\ref{causality}(a), most effort is on comparing with previous exercises and predicting the effect of adjustments.    

Related SLA models describe a student's ability to correctly change the taught knowledge in incremental steps by understanding similarities and differences from taught concepts, and predicting the effect of changes. Changes also expose the possible variability of a newly learned concept, and give the full image of a concept. Equation (\ref{eq4}) predicts the likelihood to correctly adjust a concept for a slightly different problem:
{\small
\begin{equation}
Concept~adjustment(Prob) = M_{ADJ} (SetCon, DSim,  Imp, emotion, social)
\label {eq4}
\end{equation}
}
Parameter $Prob$ is a new problem, $SetCon$ is the set of all concepts in the solution, $DSim$ is the similarity between the new exercise and previous exercises, and  $Imp$ is the student's ability to reason the effect of the modifications on the result. 
 
Modeling routines also correlate the learning rate and cognitive effort during concept adjustment with the characteristics of concept recall, like the learning rate in equation (\ref{eq3}). The cognitive process involved in adjustment includes the following dependencies among the four activities shown in Figure~\ref{causality}(b): understanding the differences of a new problem from previous examples directly suggests the associated solutions, and hence the expected results. These steps are mainly based on associative cognitive activities, like finding similarities and differences among concepts, and incremental deductive reasoning with localized structures.   
These activities depend on the learning rate and cognitive effort described using models, i.e. equation (\ref{eq3}), and also likely to be quasi-linear with the incremental changes modeled by parameter $DSim$. 

As shown in Figure~\ref{causality}(b), modeling the four activities requires learning conditional probabilities ($q_{i,j}$), e.g., depending on the amount of dissimilarity, number of required incremental changes, emotions, and social interactions. Different models are required for different dissimilarity levels. Also, the perceived progress, emotions and social interactions are likely to change attention, motivation, and confidence, which are known to be significant for learning rate \cite {Elmi_2020, Sinclair:2018, Taub:2019}. Ensembles of learned models can capture the different dissimilarity levels, emotion kinds, or interaction patterns. Correlation analysis and Principal Correlation Analysis can understand the importance of model parameters. Another option is to extract Bayesian networks \cite {Bekele2005} with nodes expressing cognitive activities (a)-(d). The likelihood of transitions between nodes is computed using information extracted during diagnostic assessment.     

\underline{\em 3. Incorrect modification}. Students must understand the causality (logic) structure of more complex programming structures, i.e. how the execution of block $C2$ determines the subsequent execution of blocks $C3$ and $C4$, followed by the evaluation of the while-condition $C1$ in Figure~\ref{sample}. Figure~\ref{causality}(c) summarizes the model. Understanding the causality requires finding all dependencies between variables, variable clusters, and the execution flow for all possible input parameters. Note that causality understanding must tackle the uncertainty due to the possible values for variables $s[i]$, $first$, and $second$. Variable~$s[i]$ determines the number of iterations, and the execution or not of the $if$~instructions to detect the searched pattern. Understanding must also cope with the huge size of the input data space, which a student addresses by picking representative values that force program execution to follow distinct scenarios. Selecting the representative values is part of understanding causality. 

Using data collected during diagnostic assessment, SLA models describe the correctness degree of a student's cognitive image about the causality of the program execution flow. Models also show the aspects that are more likely to be incorrectly understood by a student. The activities in Figure~\ref{causality}(c) are mapped to the steps in Figure~\ref{causality}(a). Identifying the causal relations between the blocks in Figure~\ref{causality}(c) can be modeled as a Bayesian network, in which nodes correspond to the structures $C_i$. 
The likelihood to correctly understand the causality for variables $count$, $first$, and $second$ is as follows:
{\small
\begin{equation}
Causal  (count) = Concept~recall (C_3 | first, second, \Psi),  Concept~recall (C_4 | first, \Psi), Concept~recall (C_3 | first, second, \Psi)
\label{eq5}
\end{equation}
}
{\small
\begin{equation}
Causal  (first) = Concept~recall (C_4 | first, s[i], \Psi),  Concept~recall (C_3 | s[i], second, \Psi), Concept~recall (C_2 | \Psi)
\label{eq5_b}
\end{equation}
}
{\small
\begin{equation}
Causal  (second) = Concept~recall (C_3 | second, s[i], \Psi),  Concept~recall (C_4 | first, s[i], \Psi), Concept~recall (C_2 | \Psi)
\label{eq5_bb}
\end{equation}
}
Parameter $\Psi$ is a student's capacity to recognize the separate execution traces of the code, hence, to identify and prioritize test data to trigger these traces (Figure~\ref{causality}(c)). Capacity is the cognitive effort used in reasoning, and the correctness of the reached conclusions. Student learning might include trial-and-error to understand how repetitively performing the steps leads to a good set of test data. Equation~(\ref{eq5}) presents the iterative process to find the execution traces:     
{\small
\begin{equation}
\Delta \Psi  = \Psi (New, Found, NewSim, CogEff, emotion, social)
\label{eq6}
\end{equation}
}
$New$ is a newly found execution trace, $Found$ are the already found traces, $NewSim$ is the similarity of $New$ compared to $Found$, and $CogEff$ is the cognitive effort to identify $New$. $\Psi$ can be modeled as a Hidden Markov Chain \cite {Feinberg2002} or as an ensemble of models conditioned by parameter $NewSim$ modulated by parameter $CogEff$. 

\underline{\em 4. Breaking down and combining}. From a ML point of view, teaching the course means traversing the learning space to expose students to the course concepts, so that student learning and course outcomes are maximized. The traversal must be conducted under the time and resource constraints of the course (i.e. teaching support). The specific traversal is uncovered (e.g., dynamically decided) depending on the progress achieved by each student. 

To complete the diagnostic assessment flow in Figure~\ref{diagnostic}, the most likely reason for an observed error is predicted by a classifier that separates the programming constructs depending on their learning rates, i.e. using equations (\ref{eq1})-(\ref{eq6}). 
{\small
\begin{equation}
~~~~~~~~~~~~~~~Likely~error = argmax~ Concept~recall, Concept~adjustment, Causal
\label{eq7}
\vspace * {-0.1in}
\end{equation}
}

{\bf 3. Automated diagnostic assessment}. The using of concepts in problem-solving can be grouped into three categories: recalling concepts, adapting / extending concepts, and modifying concepts beyond incremental changes. Misunderstandings might occur in any of the three categories. We denote $<\delta_{i,r}, \delta_{i,e}, \delta_{i,m}>$ as the misunderstandings related to recalling, extending, and modifying concept~$i$. Moreover, triplet $\{<\delta_{i,r}\}^{(p)}_i, \{\delta_{i,e}\}^{(p)}_i, \{\delta_{i,m}\}^{(p)}_i>$ describes the set of al modifications of the three kinds involving all concepts~$i$ that were used in the response produced at step~$p$, such as the student's $p$-th attempt to solve the exercise.  

\begin{figure}[t]
\centering
\includegraphics[width=6.7in]{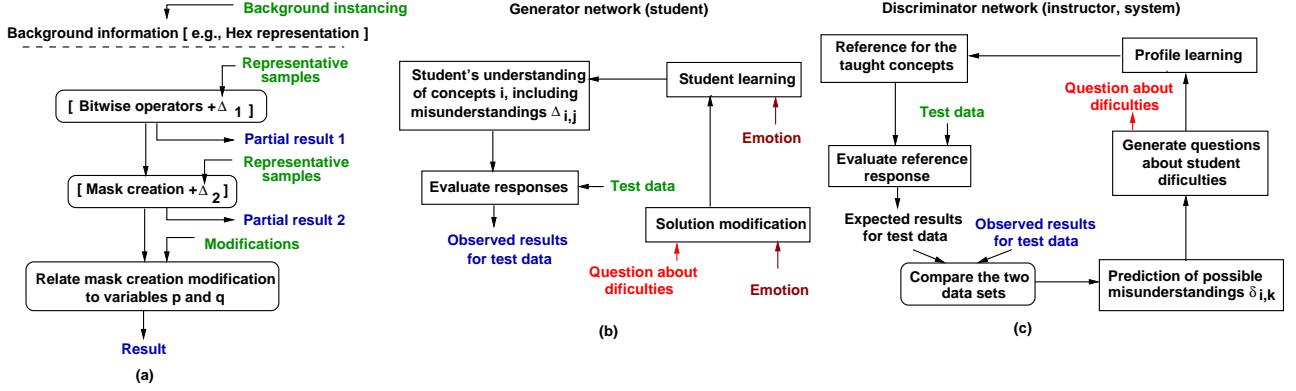}
\caption{Generator and discriminator networks for diagnostic formative assessment}
\label{fig3}
\vspace * {-0.1in}
\end{figure}

Misunderstandings produce a sequence of {\em observed results for test data}, which are the interpreted by the discriminator network. In response, the discriminator networks generate questions related the misunderstandings that must be identified. The corresponding sequence can be described as follows:
{\small
\begin{equation}
<\{\delta_{i,r}\}^{(1)}_i, \{\delta_{i,e}\}^{(1)}_i, \{\delta_{i,m}\}^{(1)}_i>, <\{\delta_{i,r}\}^{(2)}_i, \{\delta_{i,e}\}^{(2)}_i, \{\delta_{i,m}\}^{(2)}_i>
... <\{\delta_{i,r}\}^{(p)}_i, \{\delta_{i,e}\}^{(p)}_i, \{\delta_{i,m}\}^{(p)}_i> 
\label {eq8}
\end{equation}
} 

Figures~\ref{fig3}(b)-(c) illustrates the generator and discriminator networks of the implementation. The prototype of the two agent systems was recently discussed in~\cite{Doboli2020, Doboli2021}.

The generator network starts from a student's understanding of the concepts~$i$ needed to solve a certain exercise, including his / her misunderstandings $\delta_{i,j}$ of the concepts. The solutions, i.e. programs, created by a student are then evaluated for test data, selected by the student or automatically generated by the system in order to improve the covering of the variable domains. The observed results for the test data are input to the discriminator network to predict the probable misunderstandings. Based on the pointers obtained through the questions received from the discriminator network, the student modifies the exercise followed by the associated learning process. Note that the student emotion at that point is important for both Solution modification and Student learning steps. 

The generator network uses the current solution, the obtained and possible reference results, and posed questions to modify the current solution, e.g., program. Note that emotion is a parameter that influences the modification too. Solution modification can be originated by a student, or by the system in the automated mode, if the student needs cues to be able to continue the process of clarifying his / her misunderstandings. In the automated mode, the goal of the cues is not so much to show the needed modification to solve the exercise, but to guide the student towards clarifying his / her misunderstandings.   

The discriminator network uses a reference of the taught concepts, i.e. the program solution to a homework exercise. Using the same test data that was used by the generator network too, it evaluates to produce reference responses, like outputs for the executed program solution for an exercise. The two data sets are compared and their differences are then used to make predictions about possible student misunderstandings $\delta_{i,k}$. These predictions are then used to generate questions that attempt to identify the precise nature of student difficulties, e.g., their misunderstandings of the taught concepts. The generated questions become inputs to the generator network. The used test data, observed differences between reference and observed responses, expected misunderstanding $\delta_{i,k}$, and generated questions are incorporated with the rest of the data into a student's profile. 

The goal of module {\em Prediction of possible misunderstandings} is to find the sequence $\delta_{i,k}$ that converges to the sequence of desired changes $\Delta_{i,k}$, $\delta_{i,k} \xrightarrow{} \Delta_{i,k}$. The conditions for convergence are as follows, using the notation in equation~(\ref{eq8}):
{\small
\begin {equation}
\{\Delta_{i,r}\}^{(k+p)} \subseteq \{\Delta_{i,r}\}^{(k)}
\label{eq9}
\end{equation}
\begin{equation}
\{\Delta_{i,e}\}^{(k+s)} \subseteq \{\Delta_{i,e}\}^{(k)}
\end{equation}
\begin{equation}
\{\Delta_{i,m}\}^{(k+t)} \subseteq \{\Delta_{i,m}\}^{(k)}
\end{equation}
\begin{equation}
k+p, k+s, k+t \leq \max~iteration, \forall k, p, s, t 
\label{eq12}
\end{equation}
}

\begin{figure}[t]
\centering
\includegraphics[width=6.7in]{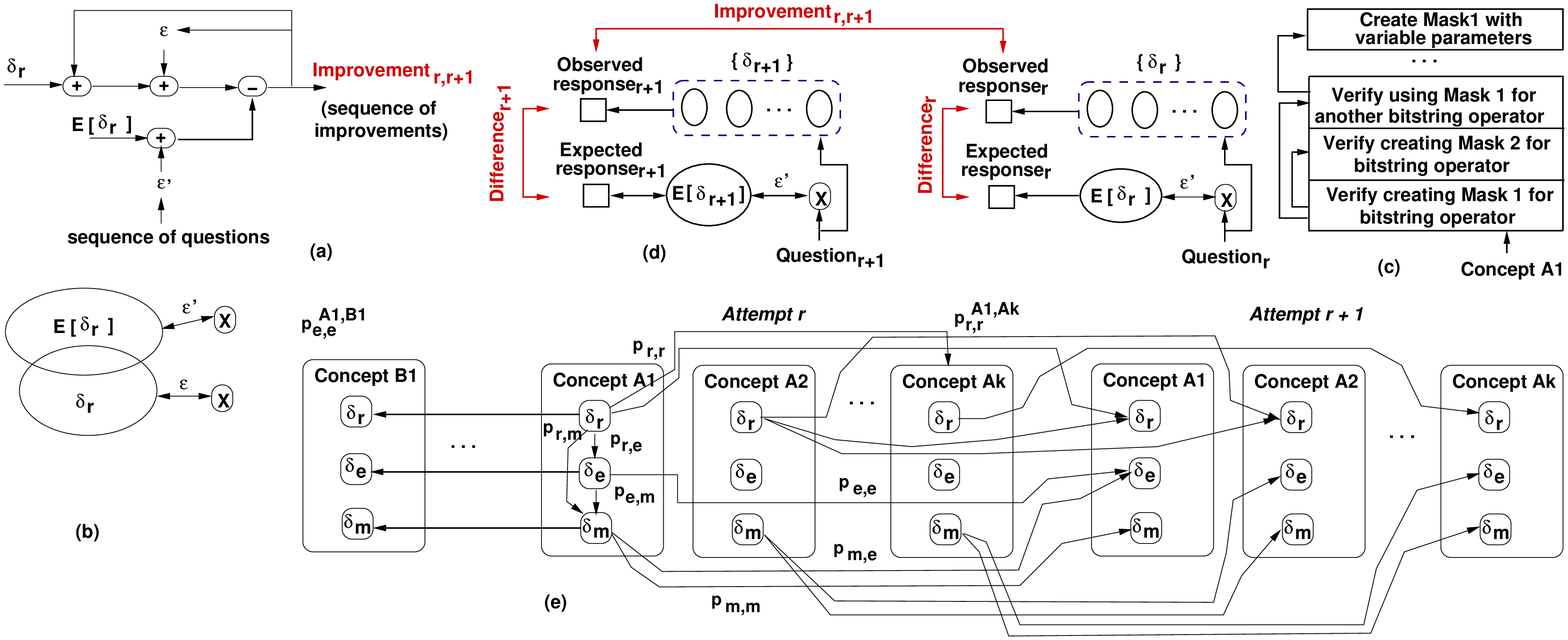}
\caption{Learning space fragment for ``Programming Fundamentals'' course}
\label{fig4}
\vspace * {-0.1in}
\end{figure}

Figure~\ref{fig4} illustrates the process that addresses equations~(\ref{eq9})-(\ref{eq12}). Figure~\ref{fig4}(a) shows the iterative procedure that adjusts the discriminator network's estimation $\delta'$ of the generator network's misunderstanding $\delta$. 
The goal of the process is to achieve two convergences: (1) $\delta' \longrightarrow \delta$, and (2) $\delta \longrightarrow \Delta$. Figure~\ref{fig4}(b) illustrates the relation between variables $\delta$ and $\delta'$ at one step of the iteration process in Figure~\ref{fig4}(a). The currently posed question attempts to address a concept or concept feature that is at distance $\epsilon'$ for the estimated misunderstanding $E[ \delta' ]$. However, in reality, the question addresses a concept or concept feature that is at a distance $\epsilon$ for the actual misunderstanding $\delta$. The convergence condition (1) is expressed by the outputs $Improvement_{r,r+1}$ in Figure~\ref{fig4}(a), which represent the difference $\delta_{r+1} - \delta_{r}$. Every question also changes the distance $\epsilon$ as well as the real misunderstanding $\delta_r$ that is tackled by the current set of questions. Note that the convergence condition (1) is achieved through questions adjusted to the image of the misunderstanding, $E[ \delta' ]$.

Figure~\ref{fig4}(d) depicts two consecutive steps $r$ and $r+1$ of the process shown in Figure~\ref{fig4}(a). Question~$r$ is posed in an attempt to clarify the expected misunderstanding $E[\delta_r]$, where the question is at a distance $\epsilon'$ from the expectation. There is a known expected response~$r$. In reality, the student's observed response~$r$ is at a difference~$r$ from the expected response. The difference can be produced by a set $\{ \delta_r \}$ of possible (actual) misunderstanding. The posed question is at a distance in set $\{ \epsilon_r \}$. A similar set of elements exists for step~$r+1$ too. The known value~$Improvement_{r+1,r}$ is the difference between the two observed responses. The prediction process attempts to find the most likely values $\delta_r$ and values $\delta_{r+1}$ that can explain the known values $Improvement_{r,r+1}$, $Difference_r$, and $Difference_{r+1}$.

The following equations describe the model in Figure~\ref{fig4}(d):
{\small
\begin{equation}
Improvement_{r,r+1} = Observed~response_{r+1} \ominus Observed~response_{r}
\label{eq13}
\end{equation}
\begin{equation}
Difference_r = Observed~response_r \ominus Expected~response_r
\end{equation}
\begin{equation}
Expected~response_r = E [ \delta_r ] \oplus \epsilon'_r (Question_r)
\label{eq15}
\end{equation}
\begin{equation}
Observed~response_r = ML~~[ \{ \delta_r \} \oplus \epsilon_r ]
\end{equation}
\begin{equation}
Difference_{r+1} = Observed~response_{r+1} \ominus Expected~response_{r+1}
\end{equation}
\begin{equation}
Expected~response_{r+1} = E [ \delta_{r+1} ] \oplus \epsilon'_{r+1} (Question_{r+1})
\label{eq17}
\end{equation}
\begin{equation}
Observed~response_{r+1} = ML~[ \{ \delta_{r+1} \} \oplus \epsilon_{r+1} ]
\label{eq19}
\end{equation}
} 

Implementing the convergence condition~(1) through repeatedly solving equations (\ref{eq13})-(\ref{eq19}) by adjusting the questions, e.g., Question~$r$ in equation~(\ref{eq15}) and Question~$r+1$ in equation~(\ref{eq17}). Figure~\ref{fig4}(c) illustrates the structuring of the question space. Questions are linked together using three links: (i) synonyms, (ii) homonyms, and (iii) higher abstraction levels. 
\begin{itemize}
\item
\vspace * {-0.15in}
{\em Synonyms} are questions that follow the same programming structure as the current or previous questions to address the same expected misunderstanding. Their purpose is to clarify misunderstandings in one's recalling of a concept or concept recalling. Figure~\ref{fig4}(c) shows questions that require solving the same problem but for other masks.

\item
\vspace * {-0.05in}
{\em Homonyms} are questions that follow a similar programming structure as the current or previous questions, but target misunderstandings on how the structure must be extended to solve new problem-solving needs. Figure~\ref{fig4}(c) presents questions that involve using masks for other kinds of bit-level operators.  

\item
\vspace * {-0.05in}
{\em Higher abstraction levels} are questions that refer to misunderstandings related to modifying and combining a recalled or extended programming structure with other previously used structures to solve a problem that diverges from the requirements of previous problems. Figure~\ref{fig4}(c) depicts questions that refer to bit level operators but using parameterized masks instead of static mask. 
\vspace * {-0.15in}
\end{itemize}

The discriminator network in Figure~\ref{fig4}(a) adjusts the sequence of questions, so that they pertain to the three categories, synonyms, homonyms, and higher abstraction levels. Figure~\ref{fig4}(e) presents the way in which the misunderstandings $\delta_r$, $\delta_e$, and $\delta_m$ corresponding to concept recalling, extension, and modification relate to each other across two successive iterations $r$ and $r+1$. Concepts $A_1$, $A_2$, ... $A_k$ are used in a programming construct of the solution, concepts $B_1$, $B_2$, ... are used in a subsequent programming construct, and so on.  

The sequence of questions ($Questions_r$) must minimize the following cost function:
{
\small
\begin{equation}
\min~[~\alpha~[~Observed~response \ominus Expected~response_r ~] + \beta~\frac{1}{\Delta_r \ominus \delta_r} ~]
\label {eq20}
\end{equation}
}
The first term captures convergence condition~(1), i.e. the observed response converges towards the expected response, and the second term captures convergence condition~(2), e.g., the misunderstandings $\delta_r$ of the involved concepts converge towards addressing the misunderstandings $\Delta_r$ pertaining to solving a programming problem.

The sequence of questions expresses the type of addressed misunderstandings, e.g., related to recall ($\delta_r$), extension ($\delta_e$), and modification ($\delta_m$), and the concepts to which it relates. In Figure~\ref{fig4}(e), parameters~$p_{i,j}^{C_s C_t}$ denote the probabilities of posing a question related to a misunderstanding of type~$j$ ($j \in \{ r, e, m \}$) related to concept $C_t$ after a question for a misunderstanding of type~$i$ ($i \in \{ r, e, m \}$) related to concept $C_s$. The following equation describes the relation between probabilities as shown in Figure~\ref{fig4}(e):  
{\small
\begin{equation}
p_{i,j}^{(r+1)~C_s C_t} = p_{j,j}^{(r)~C_t C_t} + \sum_{i \prec j} p_{i,j}^{(r)~C_t C_t} + \sum_{i, s \neq t} p_{i,j}^{(r)~C_s C_t} + \sum_{i, s \prec t} p_{i,j}^{(r+1)~C_s C_t}
\label{eq21}
\end{equation}
}   
The right side of the equation~(\ref{eq21}) includes the following four terms: The first term describes the probability that attempt~$r+1$ refers to the same misunderstanding type and the same concept as attempt~$r$. The second term indicates the probability that a preceding misunderstanding type on the same concept was clarified before tackling the current type, e.g., a recall misunderstanding was clarified before a extension misunderstanding, where both misunderstandings refer to the same concept. The third term refers to the probabilities of all misunderstanding types related to all concepts that were addressed through questions at iteration~$r$, and which relate to the misunderstanding currently tackled by questions. The last term refers to the probabilities of all misunderstanding types and concepts addressed during the current attempt to clarify, so that these concepts precede de current concept in the clarification process. 

Learning the probabilities in equation~(\ref{eq21}) to minimize equation~(\ref{eq20}) creates the model for a student's learning of a specific concept. They describe more general learning trends, like the capacity to memorize new knowledge, relate it to the already existing knowledge, knowledge restructuring if that is necessary, and problem-solving skills including the newly learned concepts. These trends are likely to change less over the immediate future. The sequence of posed questions should reflect the probabilities in equation~(\ref{eq20}). 



\begin{figure}[t]
\centering
\includegraphics[width=6.7in]{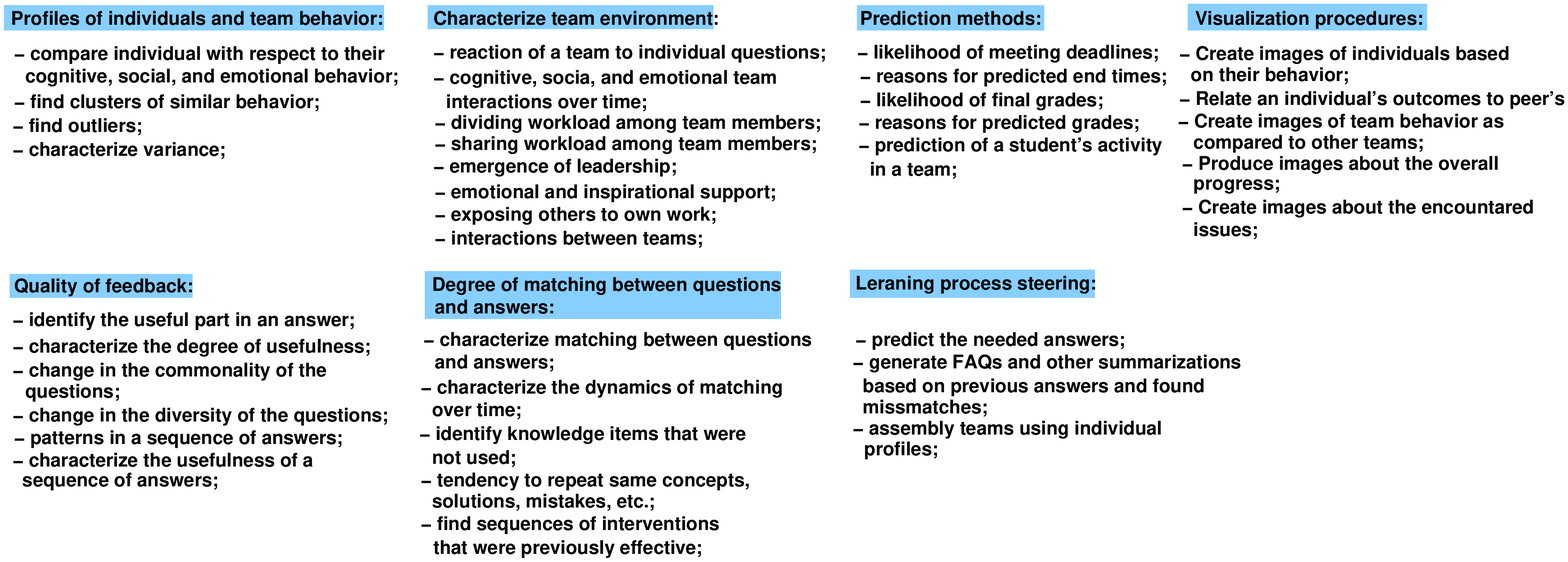}
\caption{Machine Learning library for automated diagnostic assessment}
\label{fig2}
\vspace * {-0.1in}
\end{figure}

\section {Discussions}

A main assumption of this work is that improving student learning requires customization of the educational material and delivery specific to each student background, previous learning experience, interest, priorities, available time to learn, and so on. Customization has been traditionally achieved through decreasing class sizes, increasing teaching time, and utilizing a variety of teaching modalities, e.g., lecturing, practical activities, research-based education, and so on. However, customization does not scale to increasing number of students.

Automated diagnostic assessment could ease some of the challenges to achieve scalability by focusing on understanding the specific learning activities each student struggles with. As explained in the paper, a broad set of cognitive activities pertain to learning, including correctly recalling the learned concepts from the memory, adjusting the concepts to address new needs, understanding the effects of changes on the solution, combining previous solutions, and identifying sub-goals. Continuously keeping students motivated and engaged is critical too. Moreover, the image of the presented concepts should be available to avoid biasing towards presenting certain material that is over-emphasized as opposed to material which only superficially discussed or not presented at all. 

While the implementation of the presented software system is under implementation, we have conducted a manual implementation of the supporting methodology, while developing the software module. Currently, the completed module performs team activity tracking and characterization during problem solving~\cite{Duke2021}. 

We presented next some of the surveying techniques that are being used to collect data for diagnostic assessment for each student during an entire semester. 

The following example refers to a basic programming exercise that is discussed early in the semester. The learning goal was to acquire basic programming concepts about C instructions and adjusting solution discussed in class to new but similar exercises. Data collection for assessment refers mainly to the degree to which students understood the concepts discussed in class, their capacity to correctly collect the information, and to incrementally modify the recalled information to address the exercise needs. 

{\bf Example}: Design a C program that reads from the keyboard 15 decimal values and computes (a) the average value of the positive values and (b) the average of the negative values. Display the two average values on the screen.

The data collection process for diagnostic assessment uses the following survey questions: 

(1) What steps would solve the problem? Describe in common language the solving of the exercise. (2) Identify the variables that are needed to solve the exercise. Indicate the needed variables and specify their purpose. (3) Identify the type of the variables. Indicate the type for each variable and explain your choice. (4) Read the input information from the keyword and assign the information to the related variables. Describe your understanding of the used C instructions. (5) Display the values of each variable. Explain if the displayed value was correct. If it was not, what change did you do to your program, and why? (6) How do you distinguish between positive and negative values? Explain your reasoning. (7) How do you compute the two sums and the two averages? How do you relate Step (7) to Step (6)? Explain your reasoning. (8) How do you repeat steps (4)-(7) fifteen times? Explain your reasoning. (9) Display the required output. Describe the output. If it is not, what error did you get, and how do you plan to fix it? Did your correction solve the problem? (10) How similar was the solution at Step (1) to the final solution?

The information collected from the survey is used to implement equations~(\ref{eq3}) and~(\ref{eq4}). Regarding equation~(\ref{eq3}), the collected information includes the number and type of errors and the number of attempts until correctly solving the exercise, which is then used to predict the student's learning rate and slip, the likelihood of repeating the errors, and comparisons with current peers and in previous years. Alternatively, tracking team interactions during solving the exercise permits data collection to track student emotions and social interactions. Regarding equation~(\ref{eq4}), data collection includes number of attempts until solving the exercise, the nature, place and sequence of the made changes, the prediction of the expected results for the changes, tackling multiple changes, and the situations considered during analysis and the related testing data. The constructed models predict the likelihood of a student correctly adjusting a concept in the future for a new exercise, the learning rate, the identification of uncovered situations, prediction of the situations where problems might exist in the future, and prediction of a student's capability to decompose the overall adjustments of a problem into a sequence of local adjustments.     

Early exercises also emphasized student's capabilities to use analogies and simple analogical reasoning to the exercises discussed in class to solve new problems. These exercises offer additional reinforcing of the learned concepts, but also introducing the idea of programming patterns addressing a certain type of problems. However, stressing the idea of using patterns in problem solving can also encourage a reactive response, in which students rush to use a pattern without fully understanding the specific changes of a problem. 

Assessing student capacity to modify learned concepts beyond adjustments requires insight into the degree to which they can decompose the functional requirements into sub-problems that can be mapped on the studied methods, and compose the methods while considering the causal connections between the methods. The following example illustrates this case. 

{\bf Example}: Write a C program that reads two hexadecimal values from the keyboard and then stores the two values into two variables of type unsigned char. Read two int values {\it p} and {\it n} from the keyboard, where each value is less than 8. Replace the {\it n} bits of the first variable starting at position {\it p} with the last {\it n} bits of the second variable. The rest of the bits of the first variable remain unchanged. Display the resulting value of the first variable.

The used survey assessed the capacity to decompose into sub-problems, mapping to basic methods, understanding the causal relations by predicting expected outcomes, and the linking together of the methods in the overall solution: (1) What steps would solve the problem? Describe in common language the solving of the exercise. (2) Identify the variables and their types that are needed to solve the exercise. Specify their purpose. (3) Identify the type of the variables. (3) Read the input information from the keyword and assign the information to the related variables. (4) Display the values of each variable. Explain if the displayed value was correct. If it was not, what change did you do to your program, and why? (5) What bit level operations are needed to achieve the required bit replacement? Explain your reasoning. (6) For one example, explain the required mask to solve the exercise. (7) What is the structure of the mask for any values of {\it p} and {\it n}? Explain your reasoning. (8) What algorithmic steps create the mask structure in Step (7)? Explain your reasoning. (9) Display the produced mask. Is it correct? If it is not, what error did you get, and how do you plan to fix it? Did your correction solve the error? (10) Link the code for Step (9) with the code for Step (5) that performs the bitwise operations. What interactions exist between the two code fragments? Did the resulting code work correctly? If not, what changes did you make? Did the changes solve the problem? (11) How similar was the solution at Step 1 to the final solution?

After steps (2)-(4), which define the variables and read the needed data, the solving requires decomposing the solution into two, connected parts: steps (5)-(6) for understanding the nature of the mask used in the bitwise operation and steps (7)-(9) to create the necessary mask. Step (10) focuses on the relationships that causally connect the two parts. Finally, in Step (11), students must compare their initial idea in Step (1) to the algorithm that solves the problem. 

The first part (steps (5)-(6)) requires identification of {\it p} and {\it n} values and their associated masks that originate distinct situation, like bit groups positioned at the beginning, end, or middle of the bit string. During this step, students must identify as many distinct situations as possible as well as predict the nature of the differences between the situations. 
It corresponds to equation (\ref{eq5_b}) in the model. Students use the gained insight to generalize in Step (7) how the mask structure depends on the values of {\it p} and {\it n}. It represents equation (\ref{eq5_bb}) in the model. Students should arguably form a global picture of how masks operate in any situation possible for this problem. Step (10) requires deductive reasoning to predict how the results of the first part relate to the second part. The produced mental images should include the execution flow of the code, and the relations (e.g., dependencies) between variables. It corresponds to equation (\ref{eq5}) in the model. The data collected as students repeat steps (5)-(10) while attempting to solve the exercise is used to produce equation~(\ref{eq6}). The likely type of error is predicted by equation (\ref{eq7}).

The data collected from the surveys and the corresponding models (equations (\ref{eq1})-(\ref{eq6})) are stored in a Student's Learning Aptitude (SLA).

Regarding automated cue and response generation to aid student problem solving, as shown in Figure~\ref{fig3}, the system uses student input (e.g., the devised code) to predict the nature of the errors in the code. Equation~(\ref{eq7}) predicts the nature of the error, e.g., related to recall, adjustment, and modification. Figure~\ref{fig3}(c) depicts a sequence of cues for the previous exercise (bitwise processing). Equations (\ref{eq9})-(\ref{eq12}) and (\ref{eq13})-(\ref{eq19}) is implemented as follows: a reference of the code that solves the exercise is used as a reference for automatically comparing a student's attempt to identify any errors using equation~(\ref{eq20}). 

For example, instead of definition {\tt unsigned int value} used to introduce variable {\tt value}, we noticed three kind of errors: (i) {\tt variable value}, (ii) {\tt char value}, and (iii) {\tt int value}. Note that the three errors reflect different learning limitations, hence the needed cues are different. Case~(i) describes incorrect recalling about how a variable must be defined, so a cue should suggest that any variable definition must include a type. Case~(ii) is a correct variable definition but might not work correctly if values are read using {\tt \%x} descriptor for {\tt scanf} function. Similarly, case (iii) is a correct definition and works for {\tt \%x}, but the most significant bit acts as a sign bit, thus possibly raise issues for the used bitwise operation. The cues needed for the two cases should address an incorrect adjustment of the variable definition for the needs of this exercise. Note that to some degree finding these errors is like isolating syntax, type, and semantic error during compiling, however, learning-related cues are generated instead of compiler errors. The main difference arises from the comparison of the student code and reference code, which in addition to compiler error handling, matches the variable sets and instructions of the two programs to find which is the most likely mapping between the variables of the programs.     

An example of incorrect modification occurred when a student repeatedly attempted to solve the exercise using various masks built using bitwise and ({\tt \&}) or or ({\tt |}) operations, even though the reference code use shift left and shift right operations to create the needed masks. This case reflected a situation in which difficulties were experienced on how to connect mask creation to using the mask to solve the problem. The cue should guide the student to considering other ways of creating a mask besides those already considered by the student. Once shift operations are considered, it is likely that any new errors will be either because of incorrect recall or incorrect adjustments of previously learned concepts. 

The sequence of interactions between a student solving a problem and the automated cue generation system is depicted in Figure~\ref{fig3}(e) and modeled by equation~(\ref{eq21}). Its behavior is similar to a dialog system~\cite{Doboli2020}, in which one agent is the discriminator (generates cues) and another agent the generator (generates responses to the cues). The goal of the interaction is to minimize the semantic distance between the reference code and the student's code. 

As the complexity of the taught concepts increases, it is likely that the cognitive effort needed to understand and learn them increases too, therefor maintaining the needed motivation and learning confidence of the students becomes more important. More complex concepts are expected to be harder to learn, hence a higher motivation is required to continue to allocate focus, time, and effort in learning these concepts. While addressing issues related to student motivation and confidence are critical, the current approach does not go beyond emotion tracking in team environments.

\section {Conclusions}

This paper presents a novel Cognitive - Emotional - Social (CES) and Machine Learning (ML) methodology to measure learning progress and dynamically diagnose learning shortcomings, and then to utilize the gained insight gained to optimize the learning and teaching processes. The CES-ML methodology considers dynamic diagnostic formative assessment to uncover the causes of learning shortcomings. Based on insight from cognitive psychology, the methodology groups shortcomings into four categories: recall from memory, concept adjustment, concept modification, and problem decomposition into sub-goals and sub-problems, and concept combination. Specific data models are presented to predict the occurrence of the challenges, as well as their connection into an overall model that represents a student's learning progress. In addition to cognitive aspects, the paper argues that emotional and social behavior during learning must be included too. Then, models can be used to automatically create real-time, student-specific interventions to address less understood concepts and topics. The paper discusses the using of the methodology for the dynamic assessment of an undergraduate course on programming fundamentals. We envision that the methodology will enable the creation of new adaptive pedagogical approaches to unleash undergraduate computer engineering and science student learning potential.


\end {document}